\documentclass[9pt,twocolumn]{extarticle}

\usepackage[a4paper,textwidth=504pt,textheight=636pt,columnsep=24pt,centering]{geometry}
\usepackage[T1]{fontenc}
\usepackage{mathptmx}
\usepackage[scaled=0.9]{helvet}
\usepackage{courier}
\usepackage{amsmath}
\usepackage{amssymb}
\usepackage{graphicx}
\usepackage{capt-of}

\usepackage[backend=biber,style=alphabetic,natbib=true,backref=false,maxbibnames=99,maxcitenames=2,mincitenames=1,maxalphanames=10,minalphanames=5,doi=true,isbn=false]{biblatex}
\DeclareSourcemap{\maps[datatype=bibtex]{\map{\step[fieldsource=doi,final]\step[fieldset=url,null]}}}

\let\shortcite\cite

\usepackage{xurl}
\usepackage{xcolor}
\usepackage{bm}
\usepackage[linesnumbered,ruled,vlined]{algorithm2e}
\usepackage{booktabs}
\usepackage{colortbl}
\usepackage{siunitx}
\DeclareSIUnit\revolution{rev}

\usepackage{overpic}
\usepackage{float}
\usepackage{dsfont}
\usepackage[colorlinks=true,allcolors=blue!50!black]{hyperref}

\newcommand{\norm}[1]{\left\lVert#1\right\rVert}
\newcommand{\z}{\mathbf{z}}
\newcommand{\n}{\mathbf{n}}
\newcommand{\nz}{\n_{\z}}
\newcommand{\y}{\mathbf{y}}

\newcommand{\q}{\mathbf{q}}
\newcommand{\vvec}{\mathbf{v}}
\newcommand{\uvec}{\mathbf{u}}
\newcommand{\w}{\mathbf{w}}
\newcommand{\llvec}{\mathbf{l}}
\newcommand{\p}{\mathbf{p}}
\newcommand{\f}[1]{\mathbf{f}_{#1}}
\newcommand{\ta}{\bm{\tau}}
\newcommand{\R}{\mathbb{R}}

\newcommand{\tp}{\top}

\newcommand{\Sp}{\mathbb{S}}
\newcommand{\Om}{\Omega}
\newcommand{\rhob}{\rho_b}
\newcommand{\rhof}{\rho_f}
\newcommand{\E}[1]{E_{\text{#1}}}
\newcommand{\g}{\mathbf{g}}
\newcommand{\cvec}{\mathbf{c}}

\newcommand{\slip}{\uvec_s}
\newcommand{\slipi}{\uvec_{s,i}}

\newcommand{\Pn}{\mathcal{P}_{\n_{\z}}}
\newcommand{\Pc}[1]{\mathcal{P}_{\n_{\cvec_{#1}}}}

\newcommand{\Id}{\mathrm{Id}}
\newcommand{\Kb}{\mathbf{K}_{\text{b}}}
\newcommand{\Kfl}{\mathbf{K}_{\text{f}}}
\newcommand{\K}{\mathbf{K}}
\newcommand{\pot}{\mathbf{P}_{\phi}}
\newcommand{\phitomom}{\mathbf{M}_{\phi}}

\newcommand{\pd}{p_{\textit{dyn}}}
\newcommand{\ph}{p_{\textit{hyd}}}

\newcommand{\Fdp}{\f{\textit{dp}}}

\newcommand{\Fbu}{\f{\textit{buo}}}
\newcommand{\Fgr}{\f{\textit{g}}}
\newcommand{\Fgb}{\f{\textit{gb}}}
\newcommand{\tgb}{\ta_{\textit{gb}}}
\newcommand{\tdp}{\ta_{\textit{dp}}}
\newcommand{\Fcon}{\f{\textit{const}}}

\newcommand{\tgr}{\ta_{\textit{g}}}
\newcommand{\tbu}{\ta_{\textit{buo}}}
\newcommand{\dt}{\Delta t}
\newcommand{\Ekinf}{\E{kin}}

\newcommand{\cvol}{\mathbf{c}_{\text{vol}}}
\newcommand{\cmass}{\mathbf{c}_{m}}
\newcommand{\dAz}{\, dA_{\z}}
\newcommand{\dVz}{\, dV_{\z}}

\newcommand{\dn}[1]{\partial_{\n_{#1}}}

\newcommand{\secref}[1]{Sec.~\ref{#1}}

\newcommand{\figref}[1]{Fig.~\ref{#1}}

\newcommand{\tabref}[1]{Table~\ref{#1}}

\newcommand{\eqnref}[1]{Eq.~(\ref{#1})}

\newcommand{\algref}[1]{Alg.~\ref{#1}}
\newcommand{\algreflist}[2]{Algorithms~\ref{#1} and~\ref{#2}}

\title{\vspace*{-2.5em}\textbf{Rigid Body Dynamics in Ambient Fluids}}
\author{Marcel Padilla \qquad Aviv Segall \qquad Olga Sorkine-Hornung\\[4pt]
\normalsize ETH Zurich, Switzerland}
\date{}

\hypersetup{
  pdftitle={Rigid Body Dynamics in Ambient Fluids},
  pdfauthor={Marcel Padilla, Aviv Segall, Olga Sorkine-Hornung},
}

\begin{document}

\twocolumn[{%
\csname @twocolumnfalse\endcsname
\maketitle
\vspace{-1.5em}
\begin{center}
\includegraphics[width=\textwidth]{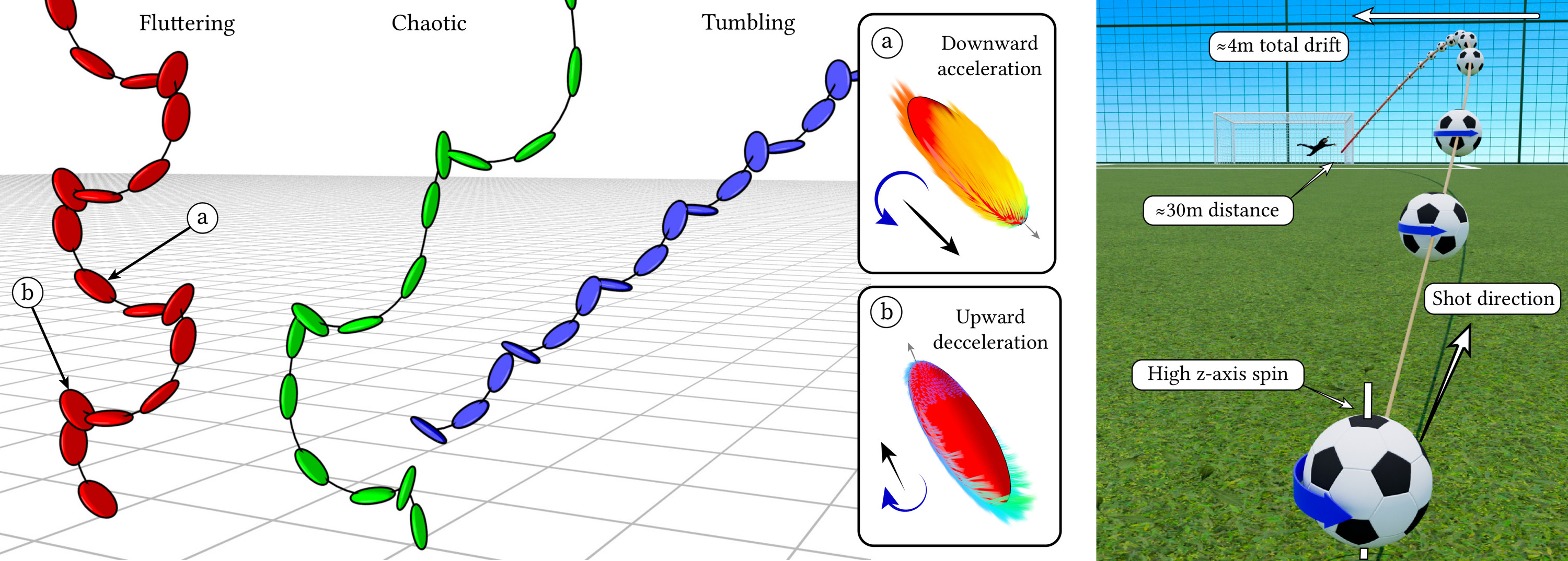}
\captionof{figure}{Our extension to rigid body dynamics enables fast simulation of fluid–body interactions without fluid solvers. We reproduce the unsteady behavior of falling plates (left) by estimating surface slip velocity and flow separation (center) to compute dynamic pressure forces. The model reproduces phenomena such as the Magnus effect, enabling curved trajectories of spinning bodies (right) as well as many other phenomena.}
\label{fig:teaser}
\end{center}
\begin{abstract}
\normalsize
We present a novel framework for rigid body dynamics in ambient media, such as air or water, enabling the reproduction of complex motion phenomena of objects without requiring computational fluid dynamics simulations.
Our method computes the added mass of the fluid and estimates dynamic pressure forces from the surface slip velocity, replacing shape-specific drag and lift coefficients with a geometry-driven pressure computation. A physically motivated flow separation criterion then estimates where the fluid detaches from the body surface, governed by a single free parameter with a clear physical interpretation.

Our method is the first within the rigid body dynamics context to reproduce the full range of falling plate behaviors: fluttering, tumbling, chaotic and steady modes, as well as phenomena such as the Magnus effect and the flight dynamics of an American football (tight spiral pass paradox).
The resulting algorithm requires only surface force evaluations at runtime and integrates seamlessly into existing physics engines for fast simulation.
\end{abstract}
\vspace{1.5em}
}]

\begin{figure*}
\centering
\includegraphics[width=\textwidth]{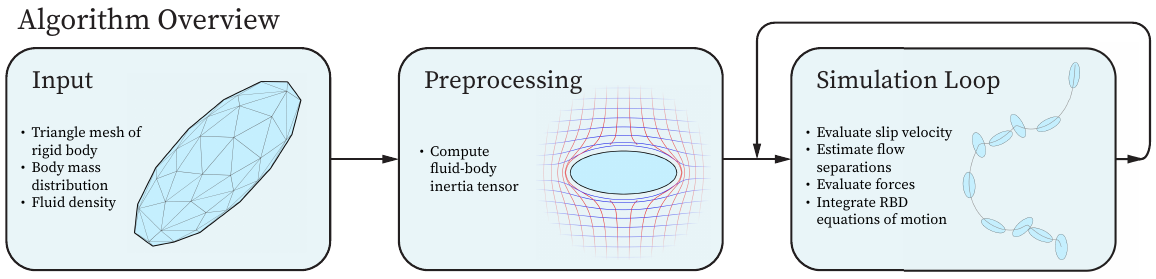}
\caption{Overview of our algorithm. Given a watertight triangulated surface mesh and physical parameters as input, we first perform preprocessing to compute quantities invariant during runtime. In the simulation loop, our proposed forces are computed and then integrated in the standard rigid body dynamics way.}
\label{fig:overview}
\end{figure*}

\section{Introduction}

Every object's motion is affected by its interaction with the \textit{ambient fluid media}, such as air or water. However, reproducing the dynamics of objects immersed in such fluids is difficult in computer graphics, either requiring computationally expensive fluid simulations or relying on simplified heuristic models that struggle to capture complex physical phenomena. Such phenomena include thin falling plates, which exhibit unstable, chaotic motion due to their interaction with air (\figref{fig:teaser}).

Traditional rigid body dynamics (RBD) frameworks effectively model motion in a vacuum and are the scalable and go-to method for the physical motion of rigid objects in modern animation software. The realism of the simulation results relies on the object's momentum being large enough for the effects of the ambient fluid to remain negligible, which is often not the case for light, small or fast-moving objects. Capturing the subtle interplay between an object and its surrounding medium, including effects such as anisotropic fluid inertia, coupled linear and angular motion, lift and drag requires a more robust approach than conventional approximations, which current tools do not provide \cite{weismann_underwater_2012,soliman_going_2024}. 

When fluid–body interactions cannot be neglected, the standard approach is to represent and simulate the surrounding fluid, incurring significant computational cost, or to adopt simplified models that approximate drag and lift forces using heuristic or data-driven coefficients (see \secref{sec:related work}).%

In this paper we present a novel framework for rigid body dynamics in incompressible ambient fluids such as air or water that enables more faithful simulations of bodies moving through fluid media without requiring volumetric fluid solvers. Our focus is on flows where large-scale turbulence is not dominant, including still or gently varying conditions. We show that in this regime, a careful treatment of added mass, flow separation and surface pressure forces suffices to reproduce a broad spectrum of complex fluid–body interaction phenomena. Our core contributions include:
\begin{enumerate}

\item \textbf{A dynamic pressure force model driven by a surface slip velocity.} Rather than prescribing drag and lift coefficients, we compute pressure forces directly from the body geometry.

\item \textbf{A flow separation criterion that estimates where the fluid detaches from the body surface}, governed by a single free parameter, the flow separation angle $\alpha$, with a clear physical interpretation and a well-motivated default. The resulting algorithm requires only surface force evaluations at runtime and integrates into standard rigid body solvers without infrastructure changes (\figref{fig:overview}).

\item \textbf{The reproduction of complex fluid–body interaction phenomena} such as fluttering, tumbling and chaotic descent, the Magnus effect, golf ball trajectories, the flight dynamics of an American football, underwater motion, buoyant motion and spinning toys.
  
\end{enumerate}

 Our approach aims to advance the modeling of fluid–body interactions and improve the efficiency vs. fidelity trade-off. The results extend rigid body dynamics to incompressible media in which large-scale turbulence is not dominant, demonstrated across the range of scenarios in \secref{sec:results}.

\section{Related work}\label{sec:related work}

The foundational equations governing the motion of rigid bodies, first formulated by Euler in the 18th century, remain central to modern mechanics \cite{landau_mechanics_1976,goldstein_classical_2002} and form the basis for many physically-based animation techniques. A widely used introduction can be found in the course notes by \citet{baraff_introduction_1997}.

\paragraph{Body–fluid interaction.}
A wide range of previous work tackles fluid-solid coupling by explicitly simulating fluid media such as air and water. The foundational techniques for this include particle-based methods \cite{carlson_rigid_2004}, direct-forcing schemes \cite{becker_direct_2009}, smoothed particle hydrodynamics formulations \cite{akinci_versatile_2012} and reduced-order models \cite{treuille_model_2006}. The reliance on full fluid simulation continues in state-of-the-art research for tasks such as control, frictional contact, stability and aerodynamic design \cite{li_difffr_2023, probst_monolithic_2023, banks_stable_2017, roenby_robust_2024, le_computational_2024}. Within such solvers, fluid separating from a solid wall is modeled by inequality (separating) boundary conditions~\cite{batty_fast_2007, chentanez_multigrid_2011}. In contrast to these approaches, our method does not require computationally expensive fluid volume simulations. Instead, it predicts the influence of the fluid on the rigid body directly via geometric arguments using only the surface of the object.

\paragraph{Motion in fluids without fluid simulations.}
Several works show that complex fluid behaviors can be captured without full volumetric fluid solvers by using physics-informed models. For example, \citet{ozgen_underwater_2010} simulate cloth dynamics using fractional derivatives that capture drag forces directly, while \citet{lannes_dynamics_2017} formulates the dynamics of floating solids purely in terms of surface pressures. Vortex-based approaches reproduce smoke, vortex rings and related effects using filament primitives instead of fluid solvers \cite{angelidis_simulation_2005,weismann_filament-based_2010,padilla_bubble_2019}. \citet{gross_motion_2023} achieve swimmer-like motions through purely shape-driven dissipation, and \citet{soliman_going_2024} replicate motion-in-fluid phenomena using simplified forces and a fast local added mass estimator on convex surfaces. \citet{weismann_underwater_2012} introduce Kirchhoff tensor-based added mass into rigid body dynamics and can be seen as the direct predecessor of our work, as further discussed in \secref{sec:fluid}. Like these methods, our work models fluid effects without representing or simulating the bulk fluid.

\paragraph{Specialized solutions.}
Several prior works build a specialized solver or force model around a particular shape class or task, typically with coefficients derived from wing or airfoil theory or calibrated to a target shape. MuJoCo~\cite{todorov2012mujoco} approximates the body with an inertia box or bounding ellipsoid and exposes five tunable coefficients covering blunt-, slender-, angular-drag and Kutta- and Magnus-lift. Design tools for free-flight model gliders~\cite{umetani_pteromys_2014} fit a data-driven aerodynamic model from measured flight trajectories, and those for bamboo-copters~\cite{nakamura_interactive_2016} prescribe wing-theory drag and lift, each tailored to its respective shape class. OmniAD~\cite{martin_omniad_2015} acquires the drag and lift of an arbitrary rigid object from a single video of its falling motion, whereas our method requires no measured data. Character-level models for bird~\cite{wu_realistic_2003} and insect~\cite{chen_real-time_2024} flight follow a similar recipe with wing-specific parameters. These methods produce desired results within their target shape and flow regime, but rely on per-shape calibration or dedicated solver machinery and do not generalize to arbitrary geometries. 

\paragraph{Simplified fluid force models.}
A more general class of methods, which we refer to as \emph{local-only} force models, applies simple drag and lift rules at each surface element from the local rigid body velocity only. Such formulations have a long history in animation~\cite{wejchert_animation_1991} and range from pure quadratic drag~\cite{pan_active_2018} to per-triangle angle-of-attack drag and lift~\cite{min_softcon_2019}. Their coefficients require empirical calibration. Closest to our work are two methods that go beyond the purely local rule by additonally modeling the fluid inertia: \citet{weismann_underwater_2012} introduce the Kirchhoff tensor into rigid body dynamics for added mass, but still pair it with a linear drag, and \citet{soliman_going_2024} combine a convex-shape Kirchhoff approximation with a per-face angle-of-attack drag and lift force (i.e.\ a local-only force component). We build on \citeauthor{weismann_underwater_2012} by adding a coefficient-free surface pressure force together with a flow separation criterion for wake regions, replacing heuristic coefficients with a geometry-aware force model that applies to arbitrary non-convex geometries.

\begin{table}[t!]
  \centering
  \caption{\label{tab:symbols}Summary of frequently used symbols.}
  \rowcolors{2}{gray!10}{white}
  \setlength{\tabcolsep}{4pt}
  \begin{tabular}{ll}
    \toprule
    Symbol & Description \\
    \midrule
    $B$ & Simply connected domain of the rigid body in $\R^3$ \\
    $\Omega$ & Exterior domain $\R^3\setminus B$ \\
    $S$ & Boundary of the rigid body (surface) $\partial B$\\
    $\z$ & Position vector in $\R^3$\\
    $\n$ & Outward unit normal to the body surface\\
    $\w$ & Rigid body angular velocity in body frame, \si{\radian\per\second}\\
    $\vvec$ & Rigid body linear velocity in body frame, \si{\meter\per\second} \\
    $\llvec$ & Rigid body angular momentum in body frame, \si{\kilo\gram\meter\squared\per\second}\\
    $\p$ & Rigid body linear momentum in body frame, \si{\kilo\gram\meter\per\second}\\
    $\rhof$ & Fluid density, \si{\kilo\gram\per\meter\cubed}\\
    $\alpha$ & Flow separation angle, \si{\radian}\\
    $\Kb$ & Body $6\times 6$ inertia tensor\\
    $\Kfl$ & Fluid $6\times 6$ inertia tensor\\
    $\K$ & Combined inertia tensor \\
    $\phi$ & Fluid velocity potential in $\Om$\\
    $(\ \ \ )^{\parallel}$ & Component tangential to the surface \\
    $(\ \ \ )^{\perp}$ & Component normal to the surface\\
    $\uvec$ & Fluid velocity field in $\Om$, \si{\meter\per\second} \\
    $\slip$ & Surface slip velocity, \si{\meter\per\second}\\
    $\Fdp$ & Dynamic pressure force, \si{\newton}\\
    $\pot$ & Potential map, $\#V \times 6$ matrix\\
    \bottomrule
  \end{tabular}
\end{table}

\paragraph{Falling modes of thin plates.}
The simulation of freely falling plates that exhibit fluttering, tumbling or chaotic motion remains an active and challenging research area dating back to Newton \shortcite{newton_newtons_1687} and Maxwell \shortcite{maxwell_particular_1890}. Numerous studies have addressed this phenomenon using full fluid simulations, often relying on Navier–Stokes solvers or immersed boundary methods to model vortex shedding and dynamic mode transitions \cite{pesavento_falling_2004,andersen_unsteady_2005,wu_flow_2015,rana_numerical_2020}. Experimental investigations have guided these results, highlighting the sensitivity of falling behavior to moments of inertia, vortex interactions and initial conditions \cite{zhong_experimental_2011,xiang_trajectory_2018,huang_experimetal_2013}. Although simplified or inviscid approaches \cite{sohn_simulation_2024} have been explored, they still rely on specialized fluid solvers or remain limited in generality (only ellipses).

In contrast to the above approaches, our method introduces a general, surface-based formulation that, for the first time, reproduces the full spectrum of falling thin-plate modes and other intricate phenomena, without requiring volumetric fluid simulation, specialized solvers, or shape-specific drag- and lift coefficients. 

\begin{figure}
\includegraphics[width=\linewidth]{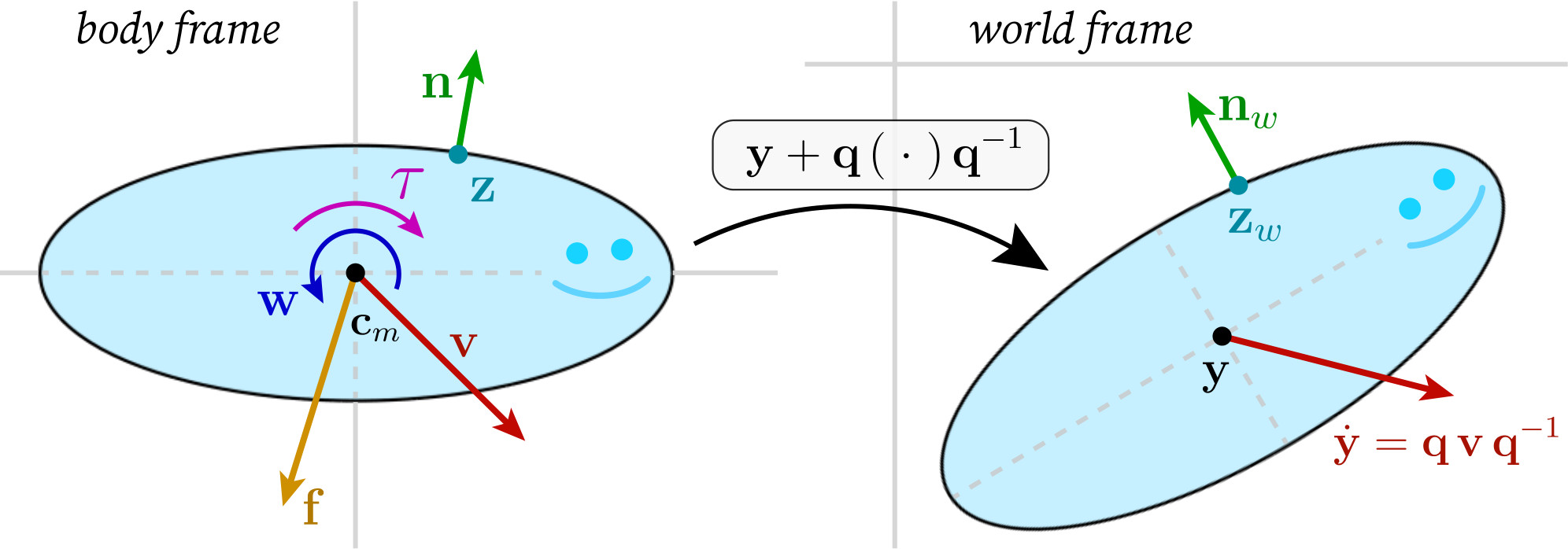}
\caption{State variables of a rigid body dynamics system. In the body frame (left), angular velocity $\w$ and linear velocity $\vvec$ define the motion, along with applied torque $\ta$ and force $\f{}$. The world frame (right) tracks the body's position $\y$ and orientation $\q \in \Sp^3$. Frame transformations are performed via quaternion conjugation, e.g., $\dot{\y} = \q \, \vvec \, \q^{-1}$.}
\label{fig:rbd}
\end{figure}

\begin{figure*}[t!]
\centering
\includegraphics[width=\linewidth]{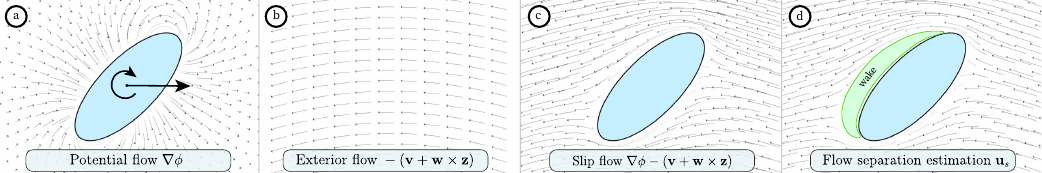}
\caption{Illustration of the relative flow around a rigid body. (a) Potential flow $\nabla \phi$ induced in the world frame by the body's motion. (b) Rigid body velocity field in the opposite direction, shown in the body frame. (c) Superposition of (a) and (b) yields the ideal potential flow past the body in the body frame. (d) Our surface slip velocity $\slip$ is the tangential part of (b) on the surface and vanishes in the wake region.}
\label{fig:flows}
\end{figure*}
\section{Background}

We briefly recall the foundational equations of rigid body dynamics (RBD), and the classical added mass of the surrounding fluid, which we subsequently extend to incorporate fluid interactions in \secref{sec:fluid}. In \tabref{tab:symbols} we provide an overview of the symbols used throughout this work.

\subsection{Rigid body dynamics}

We model a rigid body as a simply connected domain $B \subset \R^3$ enclosed by a watertight surface $S = \partial B$ with outward pointing normals $\n \in \Sp^2$. The state is described by the position $\y \in \R^3$ and orientation $\q \in \Sp^3$ (a unit quaternion, which avoids the gimbal lock singularity present in Euler angle representations), both expressed in world coordinates. The angular and linear velocities $\w, \vvec \in \R^3$ are defined in a body-fixed frame placed at the center of mass $\cmass$ (\figref{fig:rbd}). The body-fixed frame will allow us to work with a single precomputation of the potential flow solutions and transform them to the world frame. The corresponding angular and linear momenta $\llvec, \p \in \R^3$ are related to the velocities by a symmetric positive definite inertia tensor $\K \in \R^{6 \times 6}$:

\begin{equation}\label{eq:inertia}
\begin{bmatrix}
    \llvec \\
    \p
\end{bmatrix}
=
\K
\begin{bmatrix}
    \w \\
    \vvec
\end{bmatrix}.
\end{equation}
In classical RBD, which assumes a vacuum in the ambient space, $\K$ is given by the \textit{body inertia tensor} 
\begin{equation}\label{eq:body inertia}
\Kb =
\begin{bmatrix}
     \mathbf{J} & \mathbf{0} \\
    \mathbf{0} & m \, \Id\,
\end{bmatrix},
\end{equation}
where $\mathbf{J} \in \R^{3\times 3}$ is the body's rotational inertia, $m$ is its mass, and $\mathrm{Id}$ is the identity matrix. Notice how $\Kb$'s block diagonal structure decouples the angular and linear components. A position $\z \in S$ in the body-fixed frame is transformed to the world frame via $\z_w = \y + \q \ \z \ \q^{-1}$, where $\z$ is interpreted as a pure quaternion (zero scalar part). Directional quantities such as normals or forces are transformed by rotation only; the translation term $\y$ does not apply.

\subsection{Equations of motion}

The RBD differential equations are given by 
\begin{equation}\label{eq:kinematics}
    \dot{\y}  = \q \, \vvec \, \q^{-1} \ , \ \ \dot{\q} = \frac{1}{2} \, \q \, \w,
\end{equation}

\begin{equation}\label{eq:EOM}
\begin{bmatrix}
    \dot{\llvec} \\
    \dot{\mathbf{p}}
\end{bmatrix}
=
\underbrace{
\begin{bmatrix}
    \llvec \times \w + \p \times \vvec \\
    \p \times \w
\end{bmatrix}
}_{( \ \ta_{\textit{gyro}} \ , \ \f{\textit{gyro}} \ )^{\tp}:=}
+
\begin{bmatrix}
    \ta \\
    \f{}
\end{bmatrix}.
\end{equation}
The gyroscopic terms in $( \, \ta_{\textit{gyro}} \, , \, \f{\textit{gyro}} \, )$ arise because the dynamics are expressed in a non-inertial (rotating) body frame (see \citet{weismann_underwater_2012} for a deriviation). The external torque $\ta$ and force $\f{}$ in the body frame may include contributions from gravity, buoyancy or fluid interactions. Equation~\eqref{eq:inertia} is used to convert the momenta $(\llvec, \p)$ back into the angular and linear velocities $(\w, \vvec)$.

\subsection{Added mass}\label{sec:added mass}

We consider the motion of a rigid body immersed in a quiescent (i.e. still), incompressible fluid domain $\Om = \R^3\setminus B$ with constant density $\rhof$, and review the potential flow and its resulting added mass as seen in \cite{weismann_underwater_2012}. In the body-fixed frame, each point $\z \in S$ of the surface moves with the rigid body velocity $\dot{\z} = \vvec + \w \times \z$. This motion induces a velocity field $\uvec$ in the surrounding fluid, constrained by the no-penetration condition $\langle \n, \uvec \rangle = \langle \n, \dot{\z} \rangle$ on the surface. Given $\vvec$ and $\w$, the velocity field $\uvec$ that minimizes the fluid kinetic energy is the gradient of a harmonic \textit{potential} field $\phi: \Om \rightarrow \R$. This field satisfies the following Neumann boundary value problem:
\begin{equation}\label{eq:neumann-problem}
\begin{aligned}
\dn{\z} \phi(\z) &= \langle \, \n(\z), \ \vvec + \w \times \z \, \rangle,  && \text{for } \z \in S, \\
\Delta \phi(\z) &= 0, && \text{for } \z \in \Om, \\
\phi(\z) &\rightarrow 0, && \text{as } \norm{\z} \rightarrow \infty.
\end{aligned}
\end{equation}
The solution $\phi$ depends linearly on $(\w, \vvec)$, and so do the angular and linear momenta imparted to the fluid (see Appendix~\ref{app:potential flow}). This yields a linear map $\Kfl: (\w,\vvec) \mapsto (\llvec_f, \p_f)$, which we refer to as the \textit{fluid inertia tensor}. Following the classical formulation by \citet{kirchhoff_motion_1870}, we augment the rigid body’s inertia tensor to account for the influence of the surrounding fluid:
\begin{equation}
    \K = \Kb + \Kfl.
\end{equation}
This additional contribution is known as the \textit{added mass} effect, capturing the fluid’s anisotropic resistance to acceleration (Fig.~\ref{fig:flow guide}). The resulting total inertia tensor $\K$ couples the rigid body and fluid inertia into a single unified system. Unlike the decoupled block-diagonal structure of $\Kb$ in \eqnref{eq:body inertia}, the full tensor $\K$ generally includes off-diagonal terms that link angular and linear momentum.
Although $\Kfl$ is derived under potential flow assumptions, \citet{mougin_generalized_2002} and \citet{limacher_generalized_2018} show that the added-mass force is exactly additive to circulatory (vorticity-dependent) forces and independent of the Reynolds number, so that $\Kfl$ remains a valid approximation even under large-scale separation. 

\section{Fluid interactions}\label{sec:fluid}

 The fluid influences the body's dynamics primarily through two mechanisms: the transfer of momentum and the dynamic pressure along the body's surface. The potential flow enters our model only through the first mechanism, the added mass (\secref{sec:added mass}). For the second, we now introduce our core contribution: a coefficient-free, phenomenological model of the dynamic pressure force that is driven by the surface slip velocity and accounts for flow separation.
We first state our modeling assumptions (\secref{sec:flow separation}) and then define the surface slip velocity and the resulting pressure forces (\secref{sec:flow and pressure}).

\subsection{\texorpdfstring{Modeling assumptions}{Modeling assumptions}}\label{sec:flow separation}

\begin{figure}[t]
  \centering
  \setlength{\tabcolsep}{3pt}
\renewcommand{\arraystretch}{0.5}
\newlength{\cfdellgridw}\setlength{\cfdellgridw}{0.90\linewidth}
\begin{tabular}{c @{\hspace{4pt}} c}
  &
  \makebox[\cfdellgridw]{%
    \makebox[0.3333\cfdellgridw]{$0^\circ$}%
    \makebox[0.3333\cfdellgridw]{$45^\circ$}%
    \makebox[0.3333\cfdellgridw]{$90^\circ$}%
  } \\[2pt]
  \rotatebox{90}{%
    \makebox[\cfdellgridw]{%
      \makebox[0.3333\cfdellgridw]{\small Difference}%
      \makebox[0.3333\cfdellgridw]{\small Potential}%
      \makebox[0.3333\cfdellgridw]{\small CFD}%
    }%
  }
  & \includegraphics[width=\cfdellgridw]{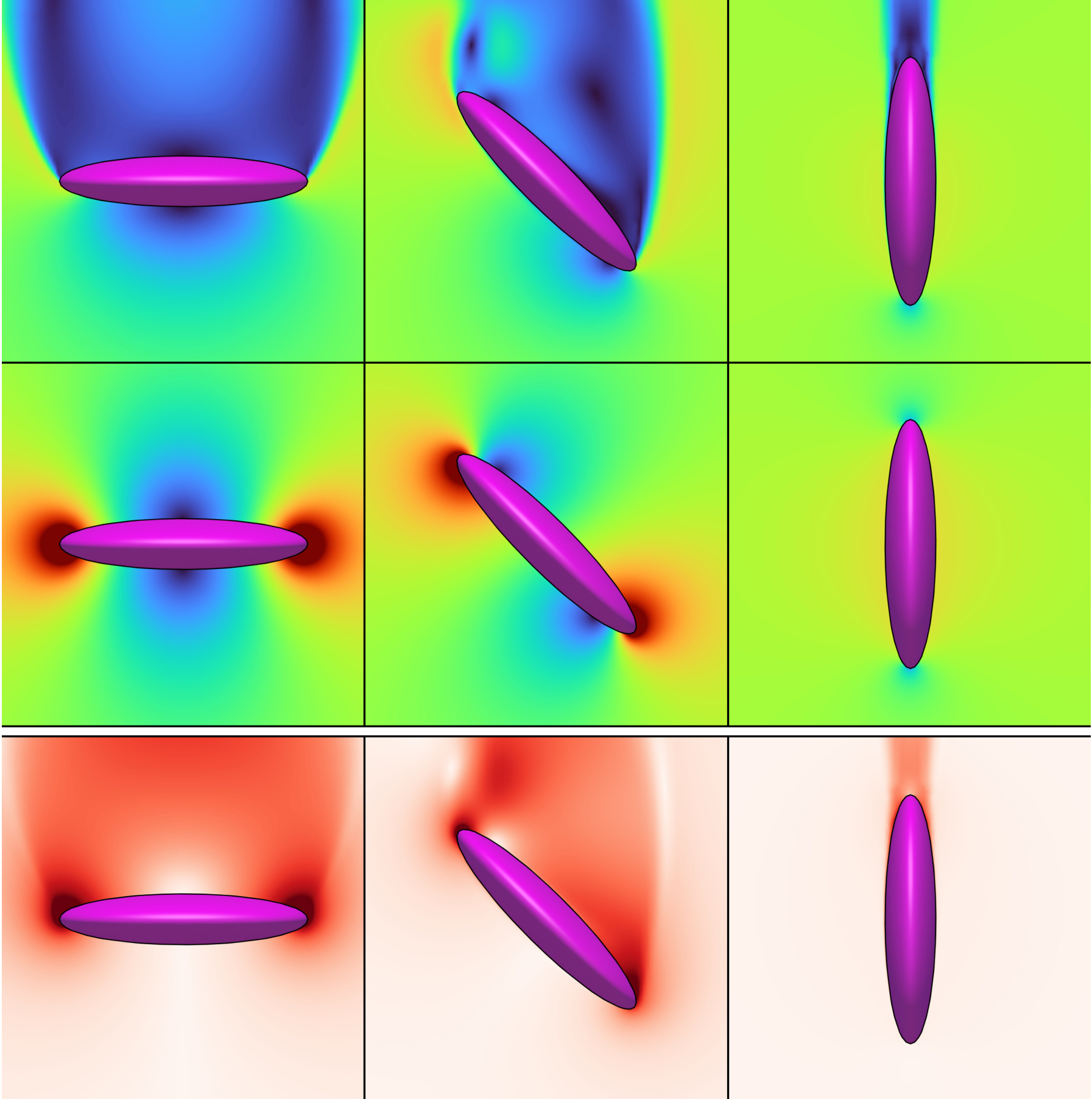} \\[4pt]
  & \includegraphics[width=\cfdellgridw]{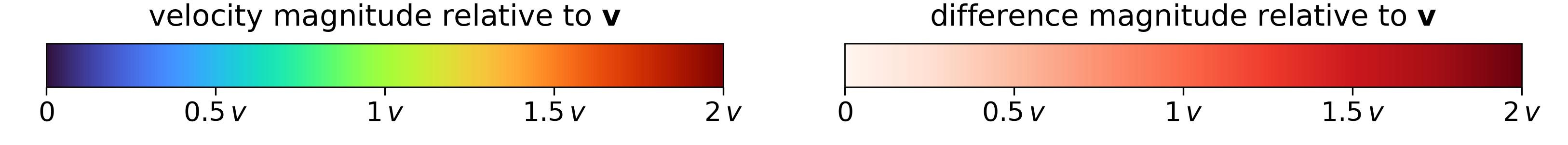} \\[4pt]
  & \includegraphics[width=\cfdellgridw]{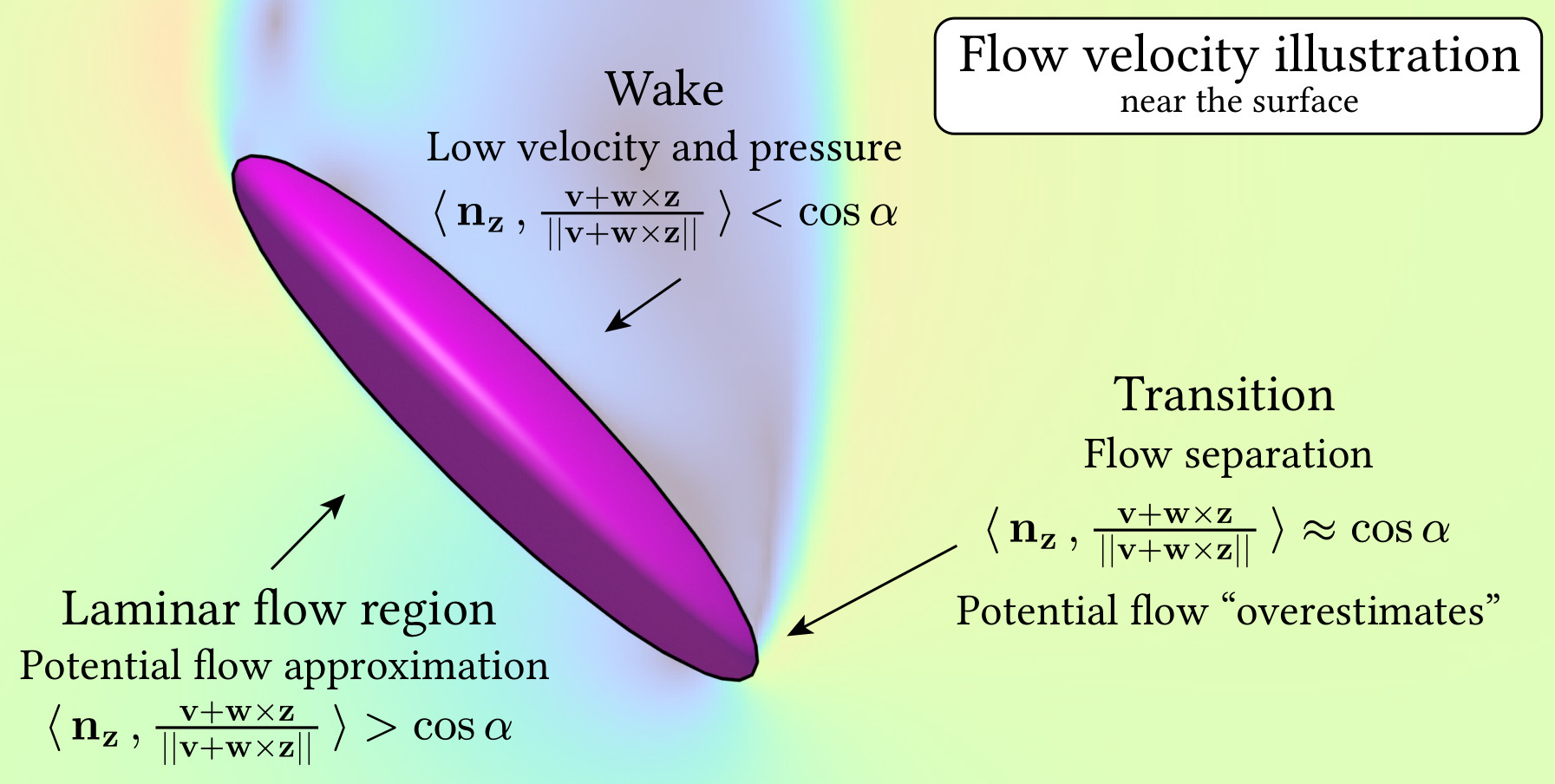} \\
\end{tabular}

  \caption{Surface velocity magnitude for an ellipsoid at three orientations ($0^\circ$, $45^\circ$, $90^\circ$) relative to the inflow, with an inflow speed of \textbf{$v = \SI{1}{\meter\per\second}$}. Each column shows the OpenFOAM~\cite{jasak2007openfoam} RANS result (top), the potential-flow estimate (middle) and their difference (bottom). The illustration (bottom) shows the surface velocity region assumptions in our model.}
  \label{fig:cfd_ellipse_variation}
\end{figure}

The fluid flows along the body surface only as long as the flow remains attached to it. At moderate to high Reynolds numbers, \textit{flow separation} typically arises along the surface, strongly affecting the surface slip velocity (\figref{fig:cfd_ellipse_variation} top).

Flow separation highlights the fundamental difficulty of \allowbreak reliably predicting fluid forces for general geometries. As a result, shape-specific drag and lift coefficients $C_D$ and $C_L$, typically parameterized by Reynolds number and angle of attack, have become standard in the literature (\secref{sec:related work}). These coefficients encapsulate pressure effects implicitly and are commonly derived through heuristic or data-driven approaches, often calibrated via wind tunnel experiments where forces are decomposed relative to a fixed global \textit{incoming flow direction}, a concept that does not transfer to bodies that also spin. Since resolving this phenomenon is highly complex and incompatible with fast simulation, we propose the following simplifying assumptions, illustrated in \figref{fig:cfd_ellipse_variation} (bottom):

\begin{description}
    \item[Assumption 1 (Flow separation).] When a point $\z\in S$ moves away from the fluid beyond a critical angular threshold $\alpha$, we estimate that the fluid flow is detached from the surface at that point.
    \item[Assumption 2 (Boundary layer).] Where the flow is attached, viscous effects in the boundary layer damp the tangential component $\nabla\phi^{\parallel}$ of the potential flow, and we omit it.
\end{description}

 Assumption~1 introduces a criterion for flow separation based on local flow attachment, used to modify the surface slip velocity. Specifically, if a surface point $ \z \in S $ is moving away from the surrounding fluid beyond a critical angle, it is considered to lie within the separated wake region, effectively shielded from external flow. In such cases, we set our surface slip velocity to zero (see \figref{fig:flows}d). This modification removes the dynamic pressure at surface points where the flow is detached.

Assumption~2 departs from the ideal potential flow past the body, $\nabla\phi - (\vvec + \w\times\z)$ (\figref{fig:flows}c). Its normal component vanishes by the Neumann condition (\eqnref{eq:neumann-problem}), but its tangential component contains $\nabla\phi^{\parallel}$, which tends to overestimate the resulting dynamic pressure. A more physically plausible model accounts for viscous effects, which dampen tangential flow near the surface due to boundary layer effects. This reduction is particularly pronounced in regions where $ \nabla \phi $ exhibits strong tangential components. We observe in our experiments that retaining $\nabla\phi^\parallel$ overestimates the dynamic pressure and suppresses unsteady behaviors such as falling-plate fluttering. This overestimation is directly visible in \figref{fig:cfd_ellipse_variation}: the potential-flow surface velocity (middle row) systematically exceeds the RANS reference (top row) near the separation boundary, where tangential flow is strongest (labeled \textit{Transition} in the illustration).

\subsection{Surface flow and pressure}\label{sec:flow and pressure}

\begin{figure}
\includegraphics[width=\linewidth]{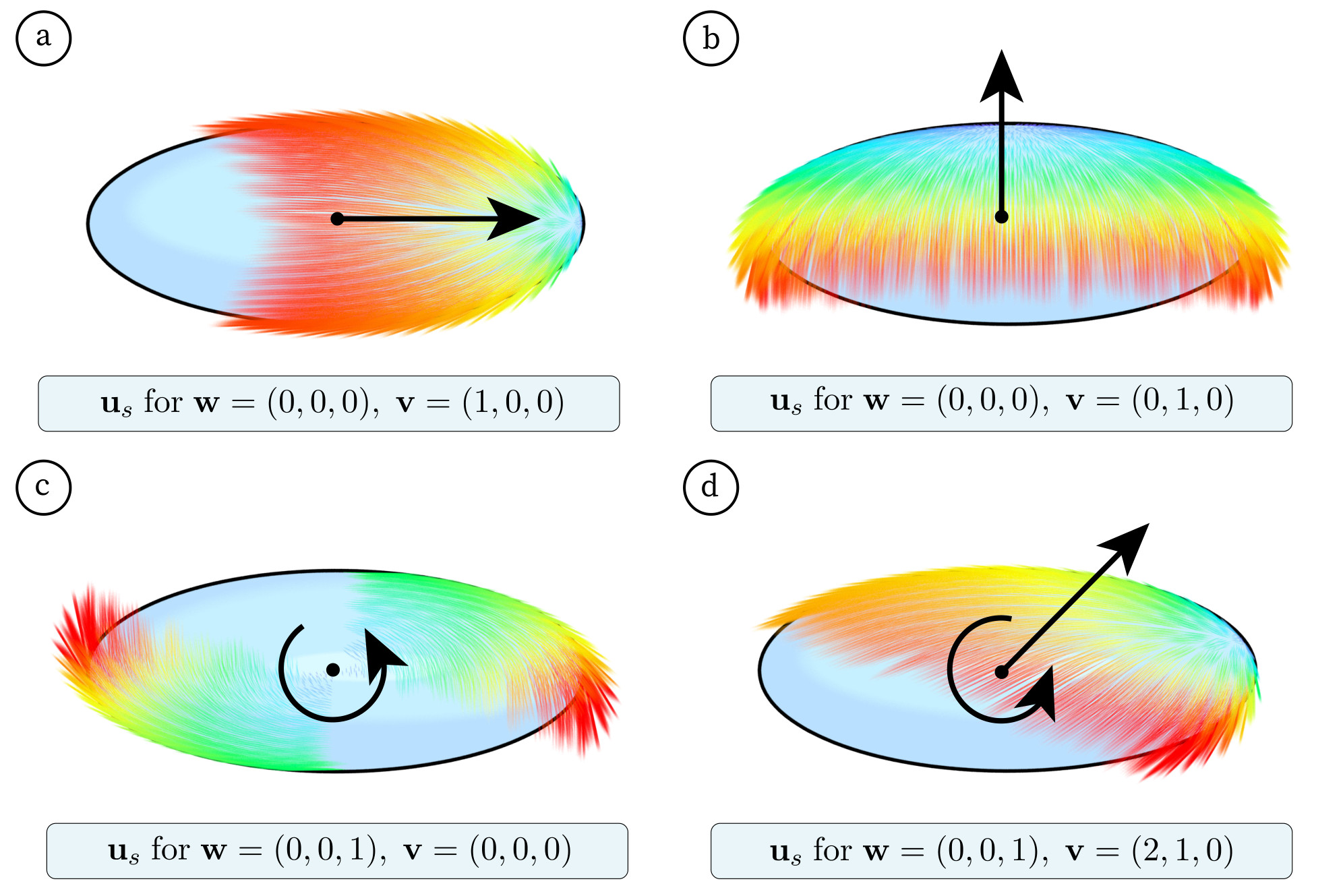}
\caption{Surface slip velocities $\slip$ (\eqnref{eq:slip}) resulting from rigid body motion $(\w, \vvec)$ for an elliptical body (3D-view). (a) Horizontal velocity. (b) Vertical velocity. (c) Rotation induces antisymmetric slip velocities. (d) A linear combination of the previous velocities. Colors indicate velocity magnitude from blue (low) to red (high). Larger slip velocities induce a larger dynamic pressure on the surface.}
\label{fig:slip figure}
\end{figure}

 We now model the pressure forces on the surface and the resulting contributions to the equations of motion. %
In the body frame, the ambient fluid moves past the surface with velocity $-(\vvec + \w\times\z)$ (\figref{fig:flows}b). Its normal component is balanced by the no-penetration condition. As we omit the tangential potential flow (Assumption~2), the surface flow is the tangential part of $-(\vvec + \w\times\z)$, obtained with the projection $\Pn(\z) := \Id - \nz\,\nz^{\top}$ onto the tangent plane, and it vanishes where the flow is detached (Assumption~1). This yields the \textit{surface slip velocity} at each point $\z \in S$:
\begin{equation}\label{eq:slip}
\slip (\z):=
\begin{cases}
 - \Pn(\z) ( \vvec + \w\times\z) & , \ \ \langle \, \n_{\z}  \,,   \tfrac{\vvec + \w \times \z}{ \norm{ \vvec + \w \times \z }} \, \rangle > \cos{\alpha}, \\
0 & , \ \ \text{else.}
\end{cases}
\end{equation}
The \textit{separation angle} $\alpha \in [\tfrac{\pi}{2},\pi]$ controls the flow detachment condition. At points where $\vvec + \w \times \z = \mathbf{0}$, both cases yield $\slip(\z) = \mathbf{0}$. Slip velocities $\slip$ with separation angle $\alpha = \frac{\pi}{2}$ are visualized in \figref{fig:slip figure}.

We model the pressure $p$ as the sum of the dynamic pressure $\pd$, hydrostatic pressure $\ph$ and a constant term $c$:
\begin{equation}\label{eq:dynamic_pressure_force}
    p(\z) = \underbrace{\tfrac{1}{2}\rhof \lVert \slip(\z)\rVert^2}_{\pd:=} + \underbrace{ \rhof \langle \z , \g \rangle}_{\ph:=} + \ \underbrace{c}_{const.},
\end{equation}
where $\g\in\R^3$ is the gravity vector in the body frame. Unlike in Bernoulli's principle, where the pressure decreases with the flow speed, our dynamic pressure $\pd$ increases with the velocity mismatch $\slip$ between the surface and the ambient fluid. We chose this form because it yields both drag and the Magnus effect. We omit the unsteady Bernoulli term $\rhof\,\partial_t\phi$: together with the quadratic term $\tfrac{1}{2}\rhof\norm{\nabla\phi}^2$, the potential-flow pressure integrates to exactly Kirchhoff's added-mass force and torque~\cite{kirchhoff_motion_1870,lamb_hydrodynamics_1895}, which \eqnref{eq:EOM} already contains through $\K$. The pressure exerts a force against the local normal direction $\nz$, resulting in the net pressure force $\f{p}$ on the rigid body, given by the dynamic pressure force $\Fdp$, the buoyant force $\Fbu$ and a constant term contribution $\Fcon$:
\begin{equation}\label{eq:pressure_force}
\begin{aligned}
    \f{p} &:= -\int_S p \, \n_{\z} \dAz \\
    &\mathrel{\phantom{:}}= \underbrace{-\int_S \pd \, \n_{\z} \dAz}_{\Fdp:=}  \underbrace{-\int_S \ph \, \n_{\z} \dAz}_{\Fbu:=}  \underbrace{-\int_S c \, \n_{\z} \dAz}_{\Fcon:=}.
\end{aligned}
\end{equation}
Using the divergence theorem, the hydrostatic pressure integrates to the familiar buoyancy, while the constant term vanishes:
\begin{equation}
 \Fbu = -\g \, \rhof\int_B1 \, dV = -\g \, \underbrace{\rhof\text{Vol}(B)}_{m_f:=} \ , \ \ \ \Fcon = -\int_{B}\nabla c \,dV=0,
\end{equation}
where $m_f$ is the mass of the fluid displaced by the body's volume. The buoyant force exerts a torque through the center of volume $\cvol\in\R^3$:
 \begin{equation}
     \tbu =  - \int_B \z\times \rhof\,\g \, \dVz = -\cvol\times m_f \, \g.
 \end{equation}

\subsection{Choice of separation angle $\alpha$}

The occurence of flow separation is often predicted by the Reynolds number $Re\in \R$, but only in a qualitative or empirical sense, and does not provide a predictive criterion to estimate a separation angle $\alpha$. $Re$ is indifferent to sharp edges that strongly facilitate flow separation, or to surface roughness, which can delay the onset of separation (an effect that is well known in the case of dimpled golf balls, see \secref{sec:golf ball}).

In practice, $\alpha$ varies little: all our examples use values between $\tfrac{\pi}{2}$ and $1.3\tfrac{\pi}{2}$. In all of our experiments (\secref{sec:results}) we denote the flow separation angle used by writing $x\tfrac{\pi}{2} = \alpha \in [\tfrac{\pi}{2},\pi]$ with $x\in [1,2]$. The choice $\alpha = \tfrac{\pi}{2}$ has a natural geometric interpretation: it designates the \textit{backside} part of the surface whose outward normal points away from the motion. This is the earliest separation our model admits: the entire downstream side of the body is treated as wake, where flow separation and base drag arise. Larger values of $\alpha$ delay the separation and shrink the wake. For a smooth sphere, laminar separation is indeed typically stated to be near $\tfrac{\pi}{2}$ from the flow direction \cite{anderson_fundamentals_2010}. We use higher values for $\alpha$ when the flow separation is expected to be delayed (e.g., golf ball or rough surfaces). We choose $\alpha$ per example by hand, guided by this expected delay and with a small amount of fine-tuning, as our separation estimate is approximate by design. The values used are listed in \tabref{tab:performance}. We note that our goal is to approximate the flow separation, and that $\alpha$ does not pinpoint a physically accurate detachment location. Instead, it estimates the pressure contribution with a single parameter.

We illustrate this separation behavior against OpenFOAM~\cite{weller1998tensorial,jasak2007openfoam} RANS simulations across a range of speeds, observing that the separated region remains consistent across Reynolds numbers, which supports using a single, fixed $\alpha$ over a large range of speeds; see Appendix~\ref{app:cfd_setup}, in particular Appendix~\ref{sec:cfd_sphere_variation}.

\section{Discretization and simulation}

We now implement the computations described in \secref{sec:fluid}. The rigid body $B$ is discretized as a triangular surface mesh with vertices and faces $(V, F)$. Each face $f_i \in F$ is associated with a barycenter $\cvec_i \in \R^3$, a unit normal $\n_i \in \R^3$ and an area $A_i > 0$. The total mass of the body and the displaced fluid are denoted by $m_b$ and $m_f$, respectively.

\begin{algorithm}[t]
\caption{\textsc{Simulation}}
\label{alg:main}
\DontPrintSemicolon
\KwIn{Triangle mesh $(V,F)$, initial state $(\q, \y, \w, \vvec)$, body mass $m_b$, fluid density $\rhof$, timestep $\Delta t$}
\KwOut{Updated states $(\q, \y, \w, \vvec)$ over time}

\textbf{Preprocessing (once per mesh):} \\
Compute $\K$ \tcp*{Inertia (\secref{sec:added mass})}

\textbf{Time integration loop:} \
\While{simulation is running}{
$\q, \y, \w, \vvec \gets$ \textsc{Integrate}$(\q, \y, \w, \vvec, \Delta t)$
}
\end{algorithm}

\begin{algorithm}[t]
\caption{\textsc{Integrate}}
\label{alg:integration}
\DontPrintSemicolon
\KwIn{State $(\q, \y, \w, \vvec)$, timestep $\Delta t$}
\KwData{Inertia $\K$, face centers $\cvec_i$, normals $\n_i$ and areas $A_i$, masses $m_b, m_f$, separation angle $\alpha$}
\KwOut{Updated state $(\q', \y', \w', \vvec')$}

$(\llvec, \p) \gets \K \, (\w, \vvec)$ \tcp*{\eqnref{eq:inertia}}

$\slipi \gets -\Pc{i}(\vvec + \w\times\cvec_i) \ \ \forall i$ \tcp*{\eqnref{eq:slip}}

$\slipi \gets \textsc{FlowSeparation}(\slipi, \alpha) \ \ \forall i$ \tcp*{\eqnref{eq:slip}}

$(\tdp, \Fdp) \gets \textsc{PressureForce}(\slip)$ \tcp*{\eqnref{eq:discrete pressure force}}

$(\ta_{\textit{gyro}}, \f{\textit{gyro}}) \gets (\llvec \times \w + \p \times \vvec, \ \p \times \w)$ \tcp*{\eqnref{eq:EOM}}

$\g \gets \q^{-1} \, \g_w \, \q$ \tcp*{Body-frame gravity}

$(\tgb, \Fgb) \gets (-\cvol \times m_f \, \g, \ (m_b - m_f) \, \g)$ \tcp*{\eqnref{eq:gravity buoyancy}}

$\ta \gets \ta_{\textit{gyro}} + \tgb + \tdp$, \ \ $\f{} \gets \f{\textit{gyro}} + \Fgb + \Fdp$ \tcp*{\eqnref{eq:EOM}}

$(\llvec', \p') \gets \textsc{Step}(\llvec, \p, \ta, \f{}, \Delta t)$ \tcp*{\eqnref{eq:EOM}}

$(\w', \vvec') \gets \K^{-1} (\llvec', \p')$ \tcp*{\eqnref{eq:inertia}}

$(\q', \y') \gets \textsc{Step}(\q, \y, \dot{\q}, \dot{\y}, \Delta t)$ \tcp*{\eqnref{eq:kinematics}}

\Return $(\q', \y', \w', \vvec')$
\end{algorithm}

\subsection{Computing the inertia}\label{sec:compute inertia}

For $\Kb$: If the body is considered to have uniform density $\rhob$, then we compute the body inertia tensor $\Kb$ according to \cite{kallay_computing_2006}. Otherwise, we prescribe masses $m_i\in\R_+$ to the barycenters of faces and compute the inertia in the classical point-based way.

For $\Kfl$: We discretize the potential $\phi$ using continuous piecewise linear hat functions centered at mesh vertices $v_i$ and, to ensure comparability and ease of implementation, solve for its values $\phi_i$ using the common \textit{potential source point method}, as done in \cite{weismann_underwater_2012}. Solving the Neumann-boundary problem (\eqnref{eq:neumann-problem}) for the $j$th component of $(\w,\vvec)$ yields the $j$th column of the $\#V \times 6$ matrix $\pot$ (see Appendix~\ref{app:potential flow}). Due to the linear hat function representation of the potential $\phi$, we can exactly realize the $6\times \#V$ matrix $\phitomom: \phi \mapsto (\llvec_f,\p_f)$ (see Appendix~\ref{app:fluid momentum}). The fluid inertia is then computed by the matrix multiplication $\Kfl=\phitomom \pot$ (Eq.~(\ref{eq:fluid inertia})). 

\subsection{Force and torque evaluations}

We evaluate the slip velocity $\slipi = \slip(\cvec_i)$ (\eqnref{eq:slip}) at each face center $\cvec_i$, including our flow separation estimation, and compute the discrete dynamic pressure force and torque using the equations:
\begin{equation}\label{eq:discrete pressure force}
\begin{aligned}
    \Fdp &= \sum_{i=1}^{\#F} -\frac{1}{2} \rhof \, \norm{ \slipi }^2 \, A_i \, \n_i, \\
    \tdp &= \sum_{i=1}^{\#F} \cvec_i \times \left( -\frac{1}{2} \rhof \, \norm{ \slipi }^2 \, A_i \, \n_i \right).
\end{aligned}
\end{equation}

To compute the gravitational and buoyant forces $\Fgr,\Fbu$ we transform the world-frame gravity vector $\g_w$ to the body frame $\g = \q^{-1} \, \g_w \,\q$. The gravitational force $\Fgr$ acts on the center of mass $\cmass$ (origin in the body frame) and therefore has vanishing torque $\tgr$. The buoyant force $\Fbu$ acts on the center of volume $\cvol$ and thus exerts a torque. We combine gravity and buoyancy into the following equations:
\begin{equation}\label{eq:gravity buoyancy}
    \Fgb := \left( m_b - m_f \right)\g \ , \ \tgb:= -\cvol \times m_f \g\,.
\end{equation}
The gyroscopic force terms $( \, \ta_{\textit{gyro}} \, , \, \f{\textit{gyro}} \, )$ of \eqnref{eq:EOM} are treated like any other force (\algref{alg:integration}).

\begin{figure}
\includegraphics[width=\linewidth]{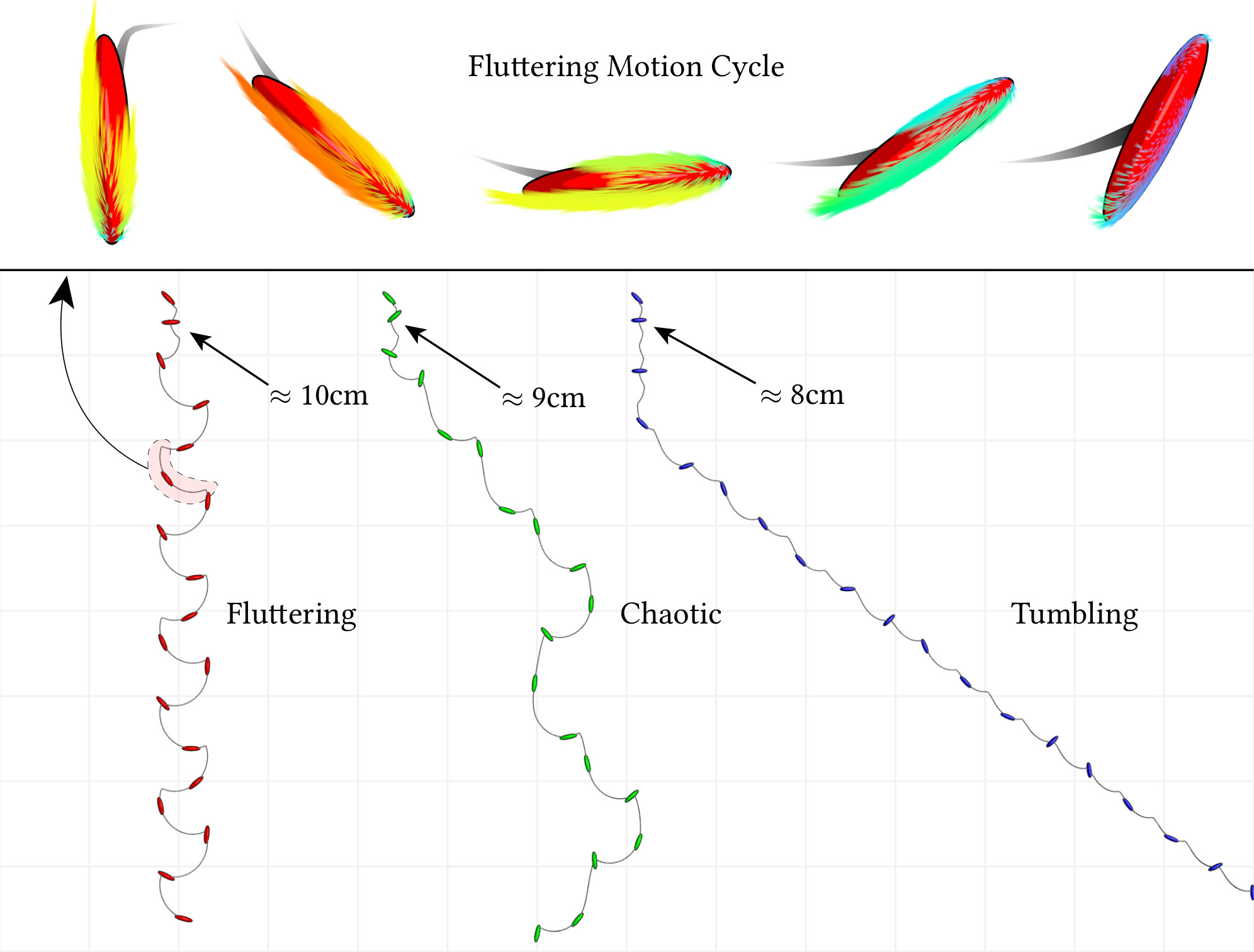}
\caption{Characteristic falling plate dynamics reproduced by our method through small variations in the width of an elliptical body. \textbf{Bottom left:} \textit{Fluttering} mode, showing periodic lateral oscillations. \textbf{Bottom middle:} \textit{Chaotic} descent, marked by irregular transitions between fluttering and tumbling. \textbf{Bottom right:} \textit{Tumbling} motion, characterized by continuous overturning. \textbf{Top:} Close-up of a fluttering cycle with surface slip velocity $\slip$ visualized using infrared-style coloring, where red indicates high slip speed and blue indicates low slip speed.}
\label{fig:falling_plates}
\end{figure}

\subsection{Algorithm and integration}

Prior to simulation, we perform a preprocessing step for each rigid body to compute the combined body–fluid inertia tensor $\K \in \R^{6 \times 6}$ (\secref{sec:compute inertia}).

Given the equations of motion (\eqnref{eq:EOM}) and the force models described here, our method supports integration using any conventional time-stepping scheme. In practice, a sympletic Euler integrator with Lie group updates for the quaternion $\q$ is sufficient in most scenarios. For cases involving large angular velocities $\w$, we adopt a fourth-order Runge–Kutta integrator with adaptive substepping to maintain stability by enforcing $\norm{ \w } \dt<$ $0.1$\,rad ($\approx 5.7^{\circ}$) per substep, a bound we chose empirically. In \algref{alg:integration}, \textsc{Step} denotes one step of the chosen time-stepping scheme.

By precomputing the fluid inertia and formulating all fluid interactions as additional forces within the RBD framework, our approach enables fast simulation of complex body–fluid interactions without full fluid simulation or heuristic lift/drag coefficients. In the following section we demonstrate that this minimal setup suffices to reproduce intricate fluid phenomena such as falling plate modes, the Magnus effect, the tight spiral pass of a football and others. The core structure of our implementation is summarized in \algreflist{alg:main}{alg:integration}.

\section{Results}\label{sec:results}

We implement our rigid body dynamics solver in Python and evaluate its performance across a range of fluid–body interaction scenarios. These experiments are designed to assess the qualitative match, generality and robustness of the proposed model. Simulation videos for all scenarios and a complete executable implementation are provided in the supplementary material.

All simulation parameters are physically determined from the respective experimental setups, with the exception of the flow separation angle $\alpha$ (\eqnref{eq:slip}), which is fixed at $\tfrac{\pi}{2}$ unless otherwise specified. The fluid density $\rhof$ is set to that of air (\SI{1.204}{\kilogram\per\cubic\meter}) or water (\SI{998}{\kilogram\per\cubic\meter}), depending on the scenario.

The fluid interaction phenomena studied in our experiments involve inherently unstable and complex dynamics, which make quantifiable ground-truth data difficult to obtain. Full fluid simulations are of limited use as a trajectory reference, since they introduce strong dependencies on the chosen numerical method and parameterization. We therefore follow the validation strategy established in prior work on rigid body dynamics in ambient fluids, focusing on reproducing known phenomena and directly comparing our results to theirs.

\subsection{Falling thin plates}

As described in \secref{sec:related work}, reproducing the unsteady dynamics of falling plates has historically relied on full fluid simulations or the use of specialized solvers. In our formulation, we model falling thin plates as light, axisymmetric ellipsoids with a width-to-height ratio of 5:1. Following the observations of \citet{andersen_unsteady_2005}, the resulting falling behavior is primarily governed by the plate’s inertia. We fix the mass at \SI{0.2}{\gram} and vary the width of the plate from \SI{8}{\centi\meter} to \SI{10}{\centi\meter} to control the moment of inertia. Our model successfully captures the characteristic falling regimes: \textit{fluttering} (a periodic lateral swaying motion) for a \SI{10}{\centi\meter} ellipse, \textit{tumbling} (continuous overturning) for an \SI{8}{\centi\meter} ellipse and the intermediate \textit{chaotic} mode (an aperiodic transition between fluttering and tumbling) for a \SI{9}{\centi\meter} ellipse. We use $\alpha = \frac{\pi}{2}$.

\begin{figure}[t]
    \centering
    \includegraphics[width=\linewidth]{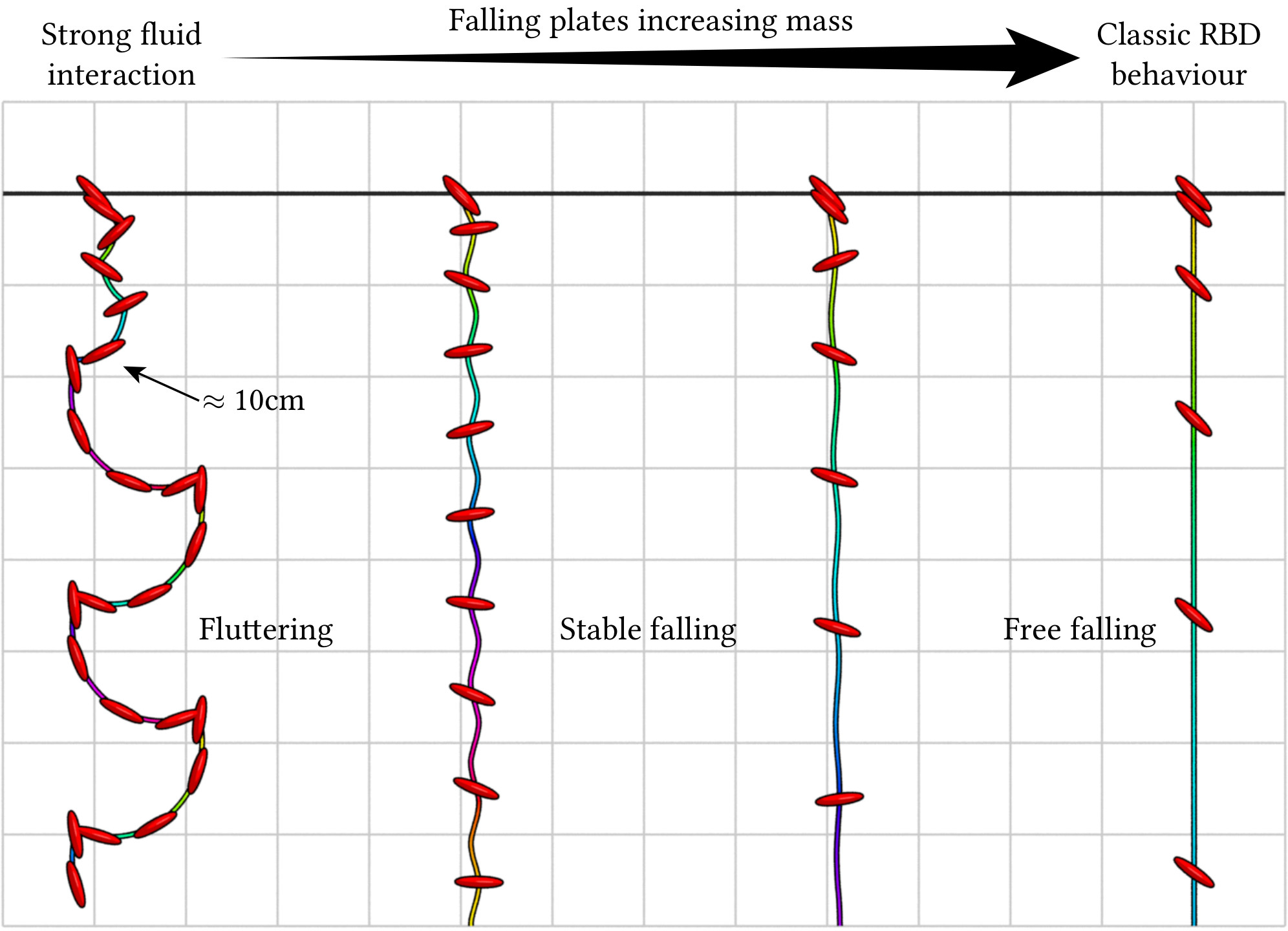}
    \caption{Falling plate trajectories for increasing mass. Light plates exhibit pronounced fluttering due to strong fluid interaction. As the mass increases, the motion transitions to stable falling (flat side tends downwards) and eventually to free fall (like in vacuum), converging toward classical rigid body dynamics. The path color indicates elapsed time through a hue cycle.}
    \label{fig:falling_plates_mass_compare}
\end{figure}

\figref{fig:falling_plates_mass_compare} illustrates the transition in the falling dynamics of thin plates as a function of mass. At low mass, pronounced fluid–body coupling induces large-amplitude fluttering trajectories. With increasing mass, the motion stabilizes into an expected stable vertical descent \cite{lau_progression_2018}. In the heavy-mass regime, fluid interactions become negligible and the trajectory converges to a free-fall motion consistent with classical rigid body dynamics, demonstrating that the method remains stable and physically consistent across the full range of mass ratios.

\begin{figure}
\includegraphics[width=\linewidth]{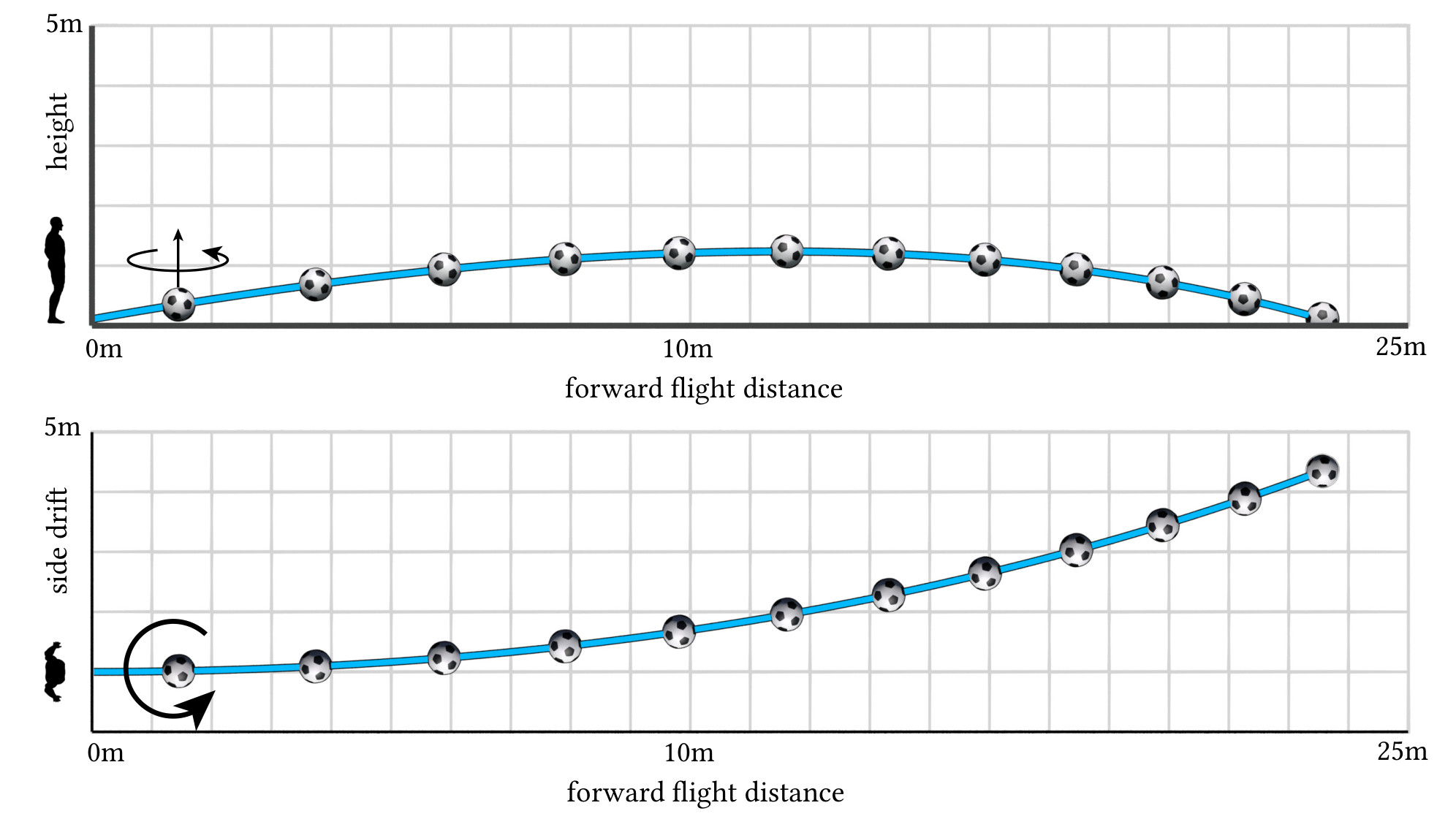}
\caption{Simulation of a spinning soccer ball exhibiting side drift due to the Magnus effect. Parameters approximate the 1997 Roberto Carlos free kick. The ball size is exaggerated for visual clarity.}    
\label{fig:soccer}
\end{figure}

\subsection{The Magnus effect}

Spinning bodies experience lateral aerodynamic forces, causing a trajectory deflection known as the Magnus effect \cite{anderson_fundamentals_2010}. We conduct three experiments that depend on this effect.

\paragraph{Curled shot of a soccer ball.}
The Magnus effect is used for \textit{curled shots}, where imparted spin causes the ball to follow a curved path. Our model naturally reproduces this effect through the relative pressure difference on each side of the spinning object.
We reproduce the iconic 1997 Roberto Carlos free kick using only estimates \cite{asal_physics_1998} of the speed ($\approx$ \SI{30}{\meter\per\second}), spin ($\approx$ \SI{10}{\revolution\per\second}), as well as the common size and weight of a soccer ball and achieve the expected \SI{4}{\meter} lateral drift (\figref{fig:teaser}, \ref{fig:soccer}). The fluid inertia $\Kfl$ plays only a minor role here: a rotating sphere displaces no fluid, and its translational added mass is half the displaced fluid mass, $\tfrac{1}{2} m_f$~\cite{lamb_hydrodynamics_1895}, i.e.\ about \SI{3.4}{\gram} for a ball of radius \SI{11}{\centi\meter}, below \SI{1}{\percent} of its \SI{430}{\gram} mass. The lateral drift is therefore produced almost entirely by the dynamic pressure force $\Fdp$.

\begin{figure}
\includegraphics[width=\linewidth]{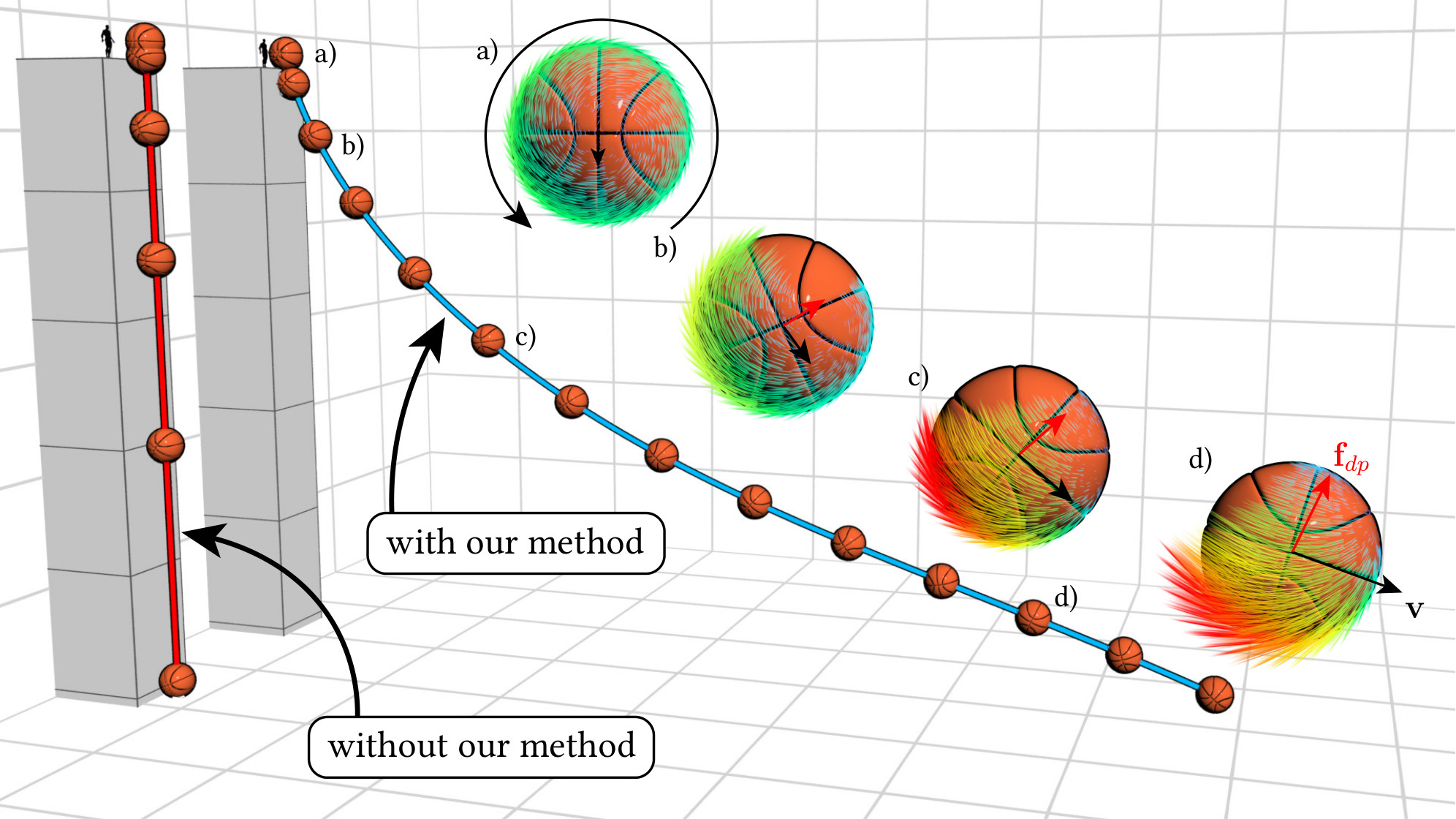}
\caption{Magnus-induced forward drift of a spinning basketball dropped from a height. The delayed onset of flow separation due to surface roughness is modeled by increasing the separation angle $\alpha=1.2\tfrac{\pi}{2}$, which delays attached flow separation and enhances lateral lift through the resulting dynamic pressure force $\Fdp$. We also show the spinning basketball trajectory using classical rigid body dynamics on the left. The ball size is exaggerated for visual clarity.}
\label{fig:basketball}
\end{figure}

\paragraph{Basketball drop.}
When a spinning basketball is dropped from a significant height, the Magnus effect induces a substantial forward drift. Using standard parameters for basketball size and mass, our model reproduces this behavior (\figref{fig:basketball}) in agreement with video footage~\shortcite{how_ridiculous_what_2015}. Due to the basketball’s rougher surface compared to a soccer ball, flow separation is delayed, which we capture by adjusting the critical separation angle to $\alpha = 1.2\tfrac{\pi}{2}$.

\subsection{Golf ball flight}\label{sec:golf ball}

Golf balls exhibit significant aerodynamic effects arising from their surface geometry and spin. Dimples on the surface induce earlier boundary layer transitions, delaying flow separation, which effectively reduces pressure drag \cite{sportsballs,choi_mechanism_2006}. In our model, this effect is represented by increasing the flow separation angle to $\alpha = 1.3\tfrac{\pi}{2}$ on a low-resolution mesh. Resolving the dimples geometrically instead would require a much denser mesh, which is computationally infeasable in our framework: the boundary integral equation (BIE) solve in our potential-flow preprocessing scales as $\mathcal{O}(\#V^3)$ (\tabref{tab:performance}).

\begin{figure}
\includegraphics[width=\linewidth]{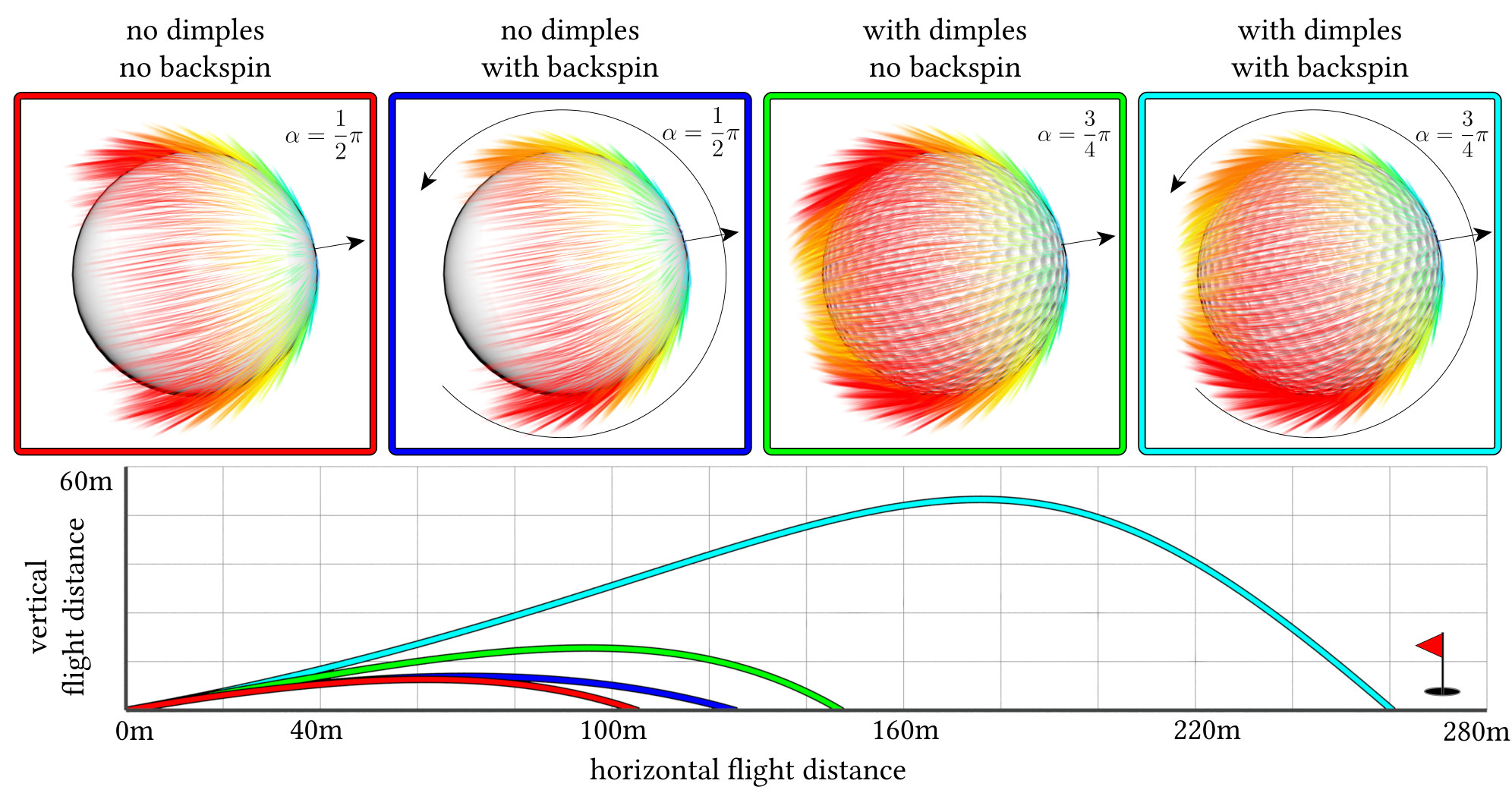}
\caption{A golf ball is shot at a launch angle of \SI{10}{\degree}, with an initial speed of \SI{78.7}{\meter\per\second}. The dimples on the ball delay the flow separation, which we model by increasing the separation angle $\alpha=1.3\frac{\pi}{2}$. Further adding a backspin (\SI{42}{\revolution\per\second}) increases the flight distance through the Magnus effect, yielding realistic flight distances as specified by the \textit{Overall Distance Standard} \protect\shortcite{usga_golf_2023, usga_randa_revised_testing_2023}. Red: a golf ball without dimples or backspin. Green: without dimples and with backspin. Blue: with dimples and without backspin. Cyan: with dimples and backspin. Top: the surface slip velocities $\slip$ are shown with infrared-style coloring, where red indicates high slip speed and blue indicates low slip speed. Bottom: the flight trajectories in a graph. Notice how the backspin causes asymmetric slip velocities and thus a stronger dynamic pressure force at the bottom of the golf ball.}
\label{fig:golf_ball}
\end{figure}

We simulate the launch of a \SI{45}{\gram} golf ball using the current test conditions of the \textit{Overall Distance Standard} \cite{usga_golf_2023, usga_randa_revised_testing_2023, usga_distance_insights_2020} (revised conditions take effect in 2028), which include a launch speed of approximately \SI{78.7}{\meter\per\second}, a launch angle of \SI{10}{\degree} and a backspin rate of \SI{42}{\revolution\per\second}. \figref{fig:golf_ball} illustrates the resulting trajectories for balls with and without dimples, as well as with and without backspin. The presence of dimples increases the effective flight range by reducing drag, while backspin generates additional lift via the Magnus effect. The simulated trajectory reaches a horizontal distance of \SI{260}{\meter}, which is within the expected range \cite{usga_golf_2023, usga_distance_insights_2020}. At these launch conditions, both dimples and backspin play a crucial role in sustaining lift and extending flight duration that our model captures. 

\subsection{Tight spiral pass}

A well-thrown American football spins rapidly around its symmetry axis, yet during flight, this axis turns to follow the curved trajectory, despite the football being symmetric in the forward-backward direction, torque-free and evenly weighted. This long-standing puzzle, known as the \textit{tight spiral paradox}, was resolved only recently~\cite{price_paradox_2020}, showing that subtle aerodynamic torques cause a precession aligning the nose with the velocity vector.

\begin{figure}
\includegraphics[width=\linewidth]{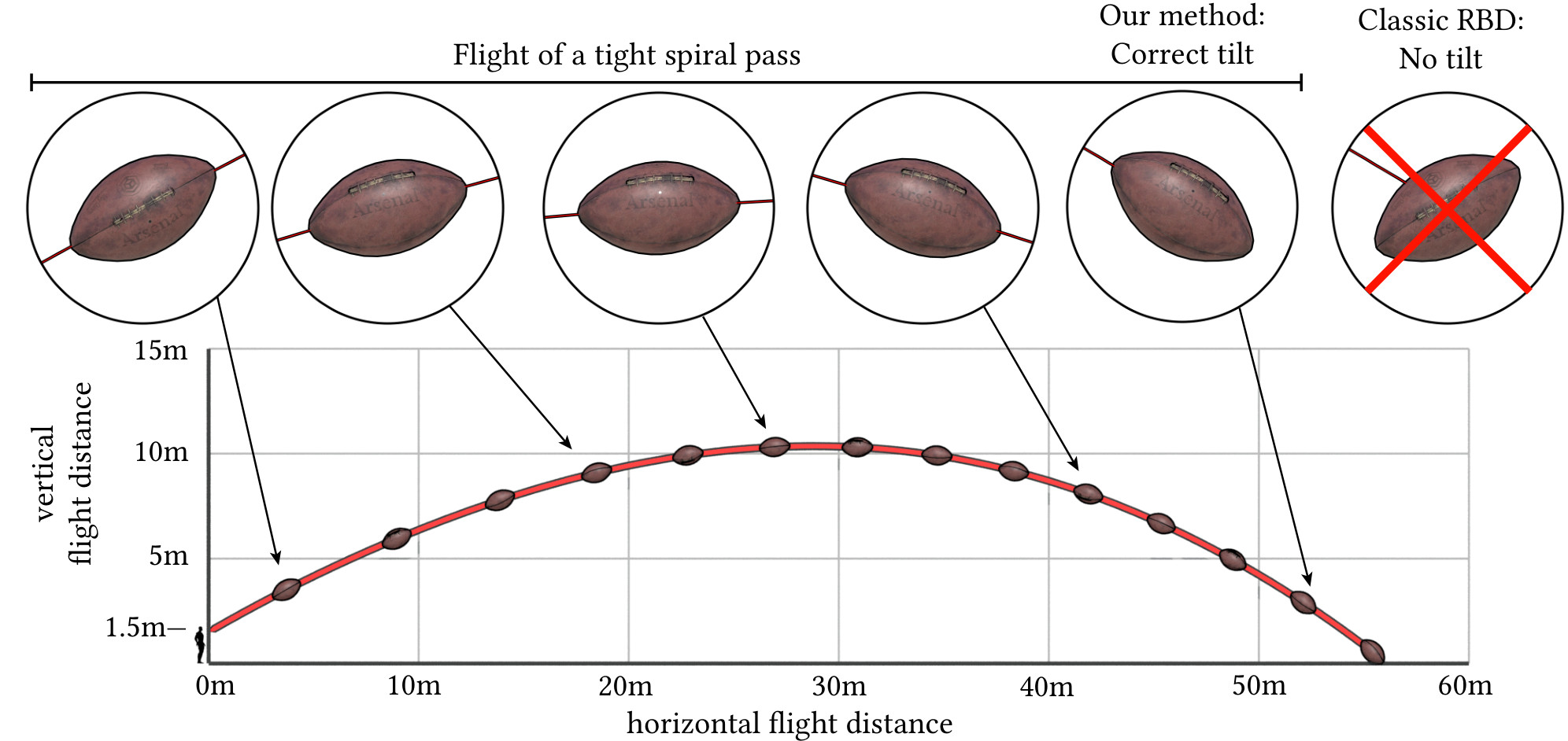}
\caption{Simulated American football throw. Our model reproduces the correct alignment of the spin axis with the trajectory, capturing the subtle aerodynamic effects responsible for the \textit{tight spiral paradox}~\cite{price_paradox_2020}. The ball size is exaggerated for visual clarity.}
\label{fig:football}
\end{figure}

We simulate a typical throw: \SI{27.4}{\meter\per\second}, \SI{10}{\revolution\per\second}, \SI{30}{\degree} angle~\cite{rae_flight_2003}, and estimate $\alpha=1.3\, \frac{\pi}{2}$ because of the rough surface and low curvature in the direction of travel reducing flow separation. When throwing the ball at an initial height of \SI{1.5}{\meter}, we recover the correct precessional alignment of the spin axis with the trajectory (\figref{fig:football}), and the ball touches the ground at approximately \SI{54.1}{\meter}. In the absence of our introduced fluid effects, the football's spin axis does not align with its trajectory (\figref{fig:football}).

\subsection{Underwater motion}

\begin{figure}
\includegraphics[width=\linewidth]{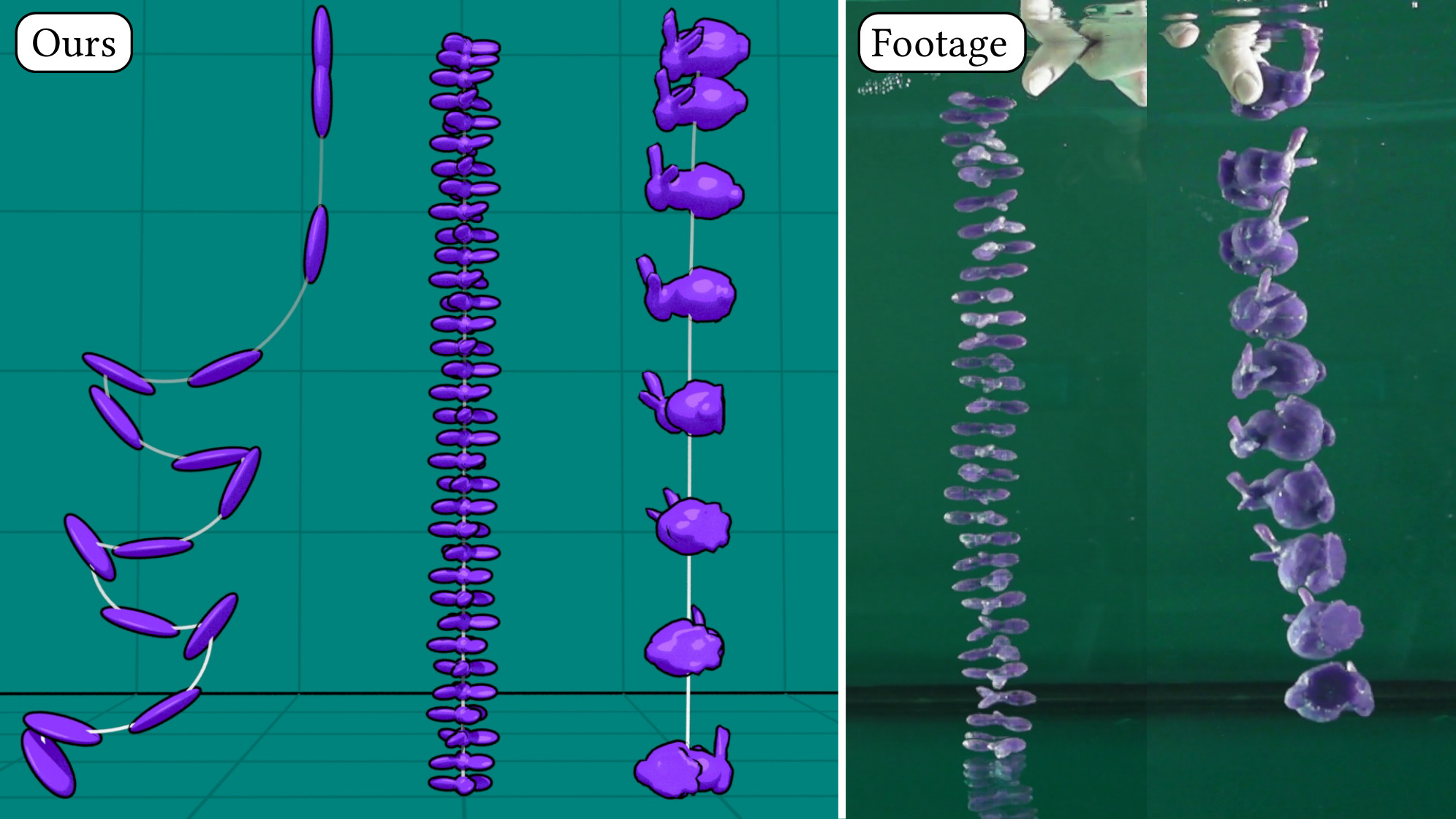}
\caption{A silicone ($\rhob =$\SI{1060}{\kilogram\per\cubic\meter}) ellipsoid, a propeller and a bunny are released underwater and compared against the experimental footage of \citet{weismann_underwater_2012}.}    
\label{fig:under_water}
\end{figure}

Simulating underwater dynamics in our model requires only adjusting the fluid density to that of water ($\rhof =$ \SI{998}{\kilogram\per\cubic\meter}), enabling direct comparison with the experimental results of \citet{weismann_underwater_2012}. We also reproduce the observed dynamics from their recordings by estimating the size and mass of the objects used in their experiments (\figref{fig:under_water}) and choosing the separation angle $\alpha$ (e.g.\ $\alpha = 1.1\tfrac{\pi}{2}$ for the bunny, \tabref{tab:performance}).

\subsection{Spinning toy propeller}

\begin{figure}
\includegraphics[width=\linewidth]{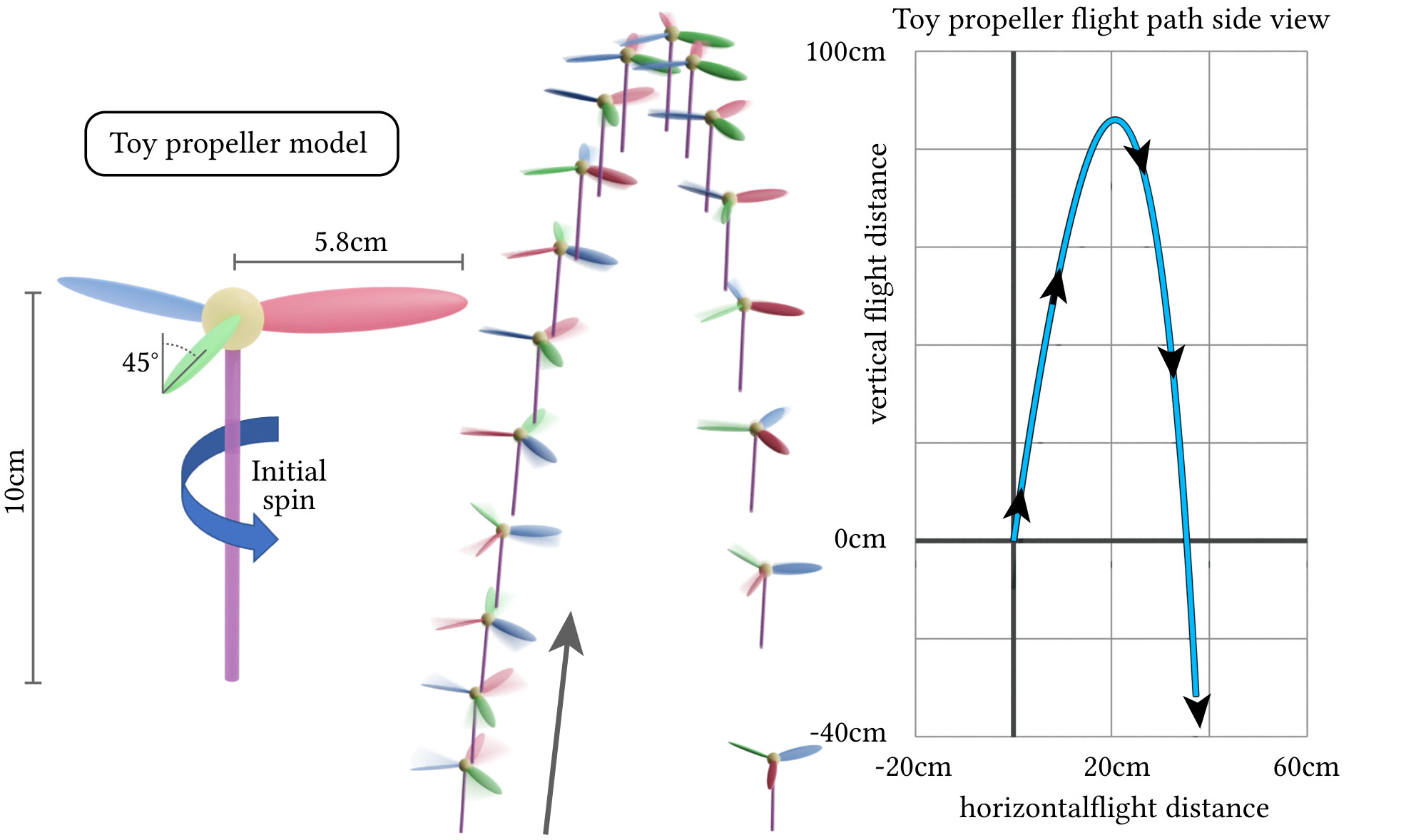}
\caption{Ascent of a spinning toy propeller reproduced by our model via aerodynamic lift generated by a high initial angular momentum. The propeller starts slightly tilted with zero linear momentum $\p$ and takes off by converting angular momentum into linear momentum due to the fluid interaction.}    
\label{fig:toy_propeller}
\end{figure}

When rapidly spun and released, lightweight propeller toys convert rotational motion into lift through aerodynamic interactions \cite{nakamura_interactive_2016}. Our model reproduces this ascent behavior without requiring any additional modifications, as shown in \figref{fig:toy_propeller}. We tilt the propeller slightly to achieve a curved flight path. We use $\alpha = \frac{\pi}{2}$.

\subsection{Buoyant dynamics of balloons}

\begin{figure}
\includegraphics[width=\linewidth]{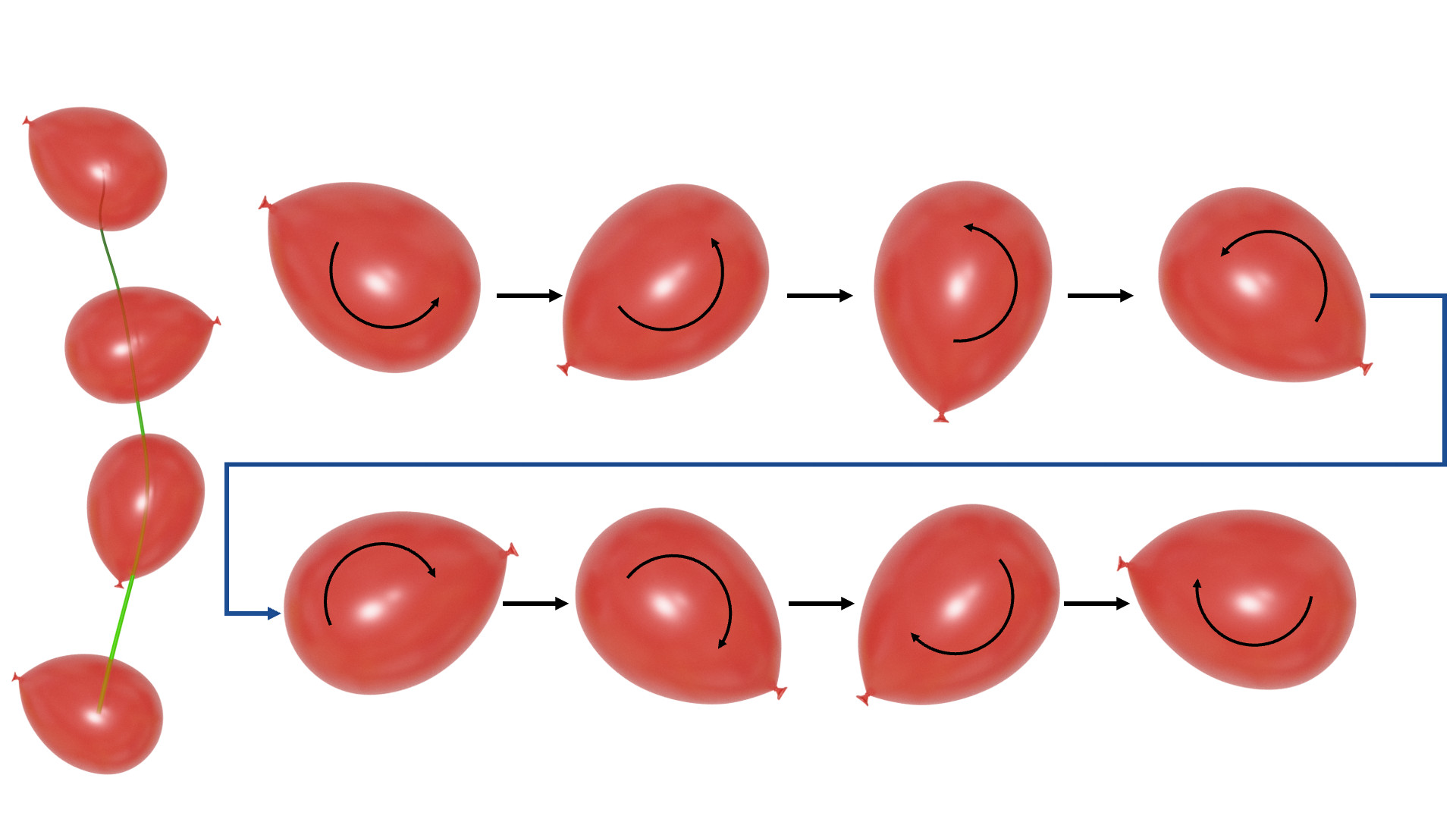}
\caption{Simulated buoyant motion of an air balloon. The offset between center of mass and center of buoyancy induces a destabilizing torque, leading to unsteady, realistic dynamics. The balloon is initially at rest.}
\label{fig:balloon}
\end{figure}

Balloons exhibit highly unsteady motion due to the large torque induced by the offset between their center of mass and center of volume. This buoyant torque leads to pronounced rotational instability even under small perturbations. We simulate a common air-filled party balloon about \SI{28}{\centi\meter} in size with a total mass of \SI{13}{\gram}, which includes the enclosed air, so that it weighs only slightly more than the air it displaces. Released from rest, it falls slowly while rocking and drifting sideways, as seen in real footage (see supplementary video). Our model realizes the horizontal motion ($\alpha = \frac{\pi}{2}$, \figref{fig:balloon}) where comparison methods fail (\figref{fig:balloon_comparison}).

\subsection{Spinning paper copter}
\begin{figure}[t]
    \centering
    \includegraphics[width=\linewidth]{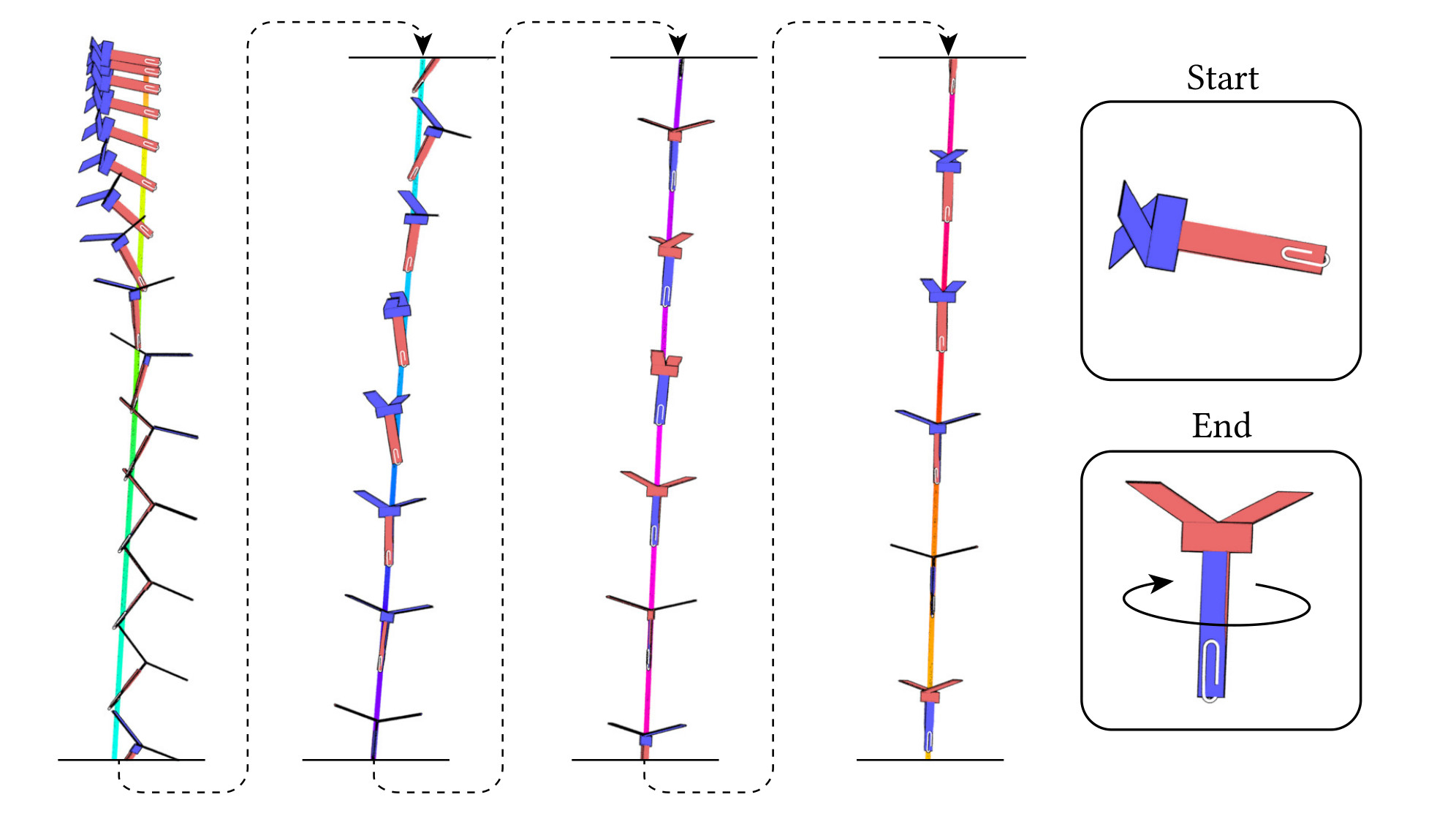}
    \caption{Simulation of a freely falling paper copter exhibiting autorotation. The offset mass distribution induces a restoring torque that aligns the copter with its descent direction, leading to stable autorotation.}
    \label{fig:paper_copter}
\end{figure}

To further demonstrate the generality of our approach, we simulate the free fall of a light paper copter. The offset mass distribution and dynamic pressure drag generate a restoring torque that passively aligns the copter’s axis with the direction of descent. This alignment promotes autorotation, stabilizing the fall through the surrounding fluid (\figref{fig:paper_copter}, $\alpha = \frac{\pi}{2}$). We regard this example as a limitation, as the thin structure had to be replaced with a slightly wider proxy geometry due to the difficulty of solving the Neumann boundary problem at thin edges (\secref{sec:limitation}). Accordingly, we interpret this result as a qualitative demonstration of autorotation rather than as a quantitative or frame-by-frame comparison to a real paper copter, unlike \citet{soliman_going_2024}, who report a close match to real-footage descent of a paper copter. 

\subsection{Comparison with existing work}\label{sec:comparison}

We evaluate our algorithm in direct comparison with the two most closely related approaches that address rigid body dynamics in ambient fluids and explicitly incorporate the fluid inertia tensor in the equations of motion. In addition, we compare against a representative local-only force model that uses no added mass. Following the validation strategy established in earlier work, we focus on reproducing known phenomena and include classical rigid body dynamics as a baseline reference.

\textit{Underwater rigid body dynamics (UWRBD) \cite{weismann_underwater_2012}.} 
 The UWRBD work employs the same computation for the inertia tensor $\Kfl$ as we do, but introduces an additional scaling factor $\lambda$ to artificially \emph{balance angular and linear inertia}. While this can improve certain outcomes, the factor has no physical basis (we set $\lambda=1$). The method further relies on a linear drag model with a free parameter that controls stability. We apply it per face as $\mathbf{f}_i = c\,\big(\nabla\phi^{\parallel} - \Pc{i}(\vvec + \w\times\cvec_i)\big)$ at each face center $\cvec_i$, i.e.\ proportional to the ideal potential flow past the body without flow separation (\secref{sec:flow separation}), with $c$ in \si{\kilo\gram\per\second}: $c = 5\cdot 10^{-7}$ for the falling plates, the toy propeller and the balloon, $10^{-5}$ for the golf ball, and $3\cdot 10^{-4}$ for the soccer ball.

\textit{Going with the flow (GwtF)\cite{soliman_going_2024}.} 
GwtF approximates the inertia tensor with a formulation that is efficient to compute but valid only for convex shapes. Their force formulation combines lift and drag coefficients into a per-face rule, fitted once to flat-plate measurements (the coefficients are given below), so it has no free parameter that we tune. Although their primary goal is to analyze the momentum evolution of deforming shapes, applying their model to a fixed geometry yields a rigid body dynamics simulation.

\textit{Local-only models (Local).}
We compare against a representative model of the common approach of computing fluid forces using only the local triangle velocity and angle of attack without any added mass. There are many possible choices of lift and drag coefficients, and we choose the same ones as GwtF for consistency, which means $C_D = 2\sin^2 (\beta), C_L = \sin (2\beta)$, where $\beta$ is the angle of attack. It differs from GwtF due to the absence of the fluid inertia.

\begin{figure}[t]
    \centering
    \includegraphics[width=\linewidth]{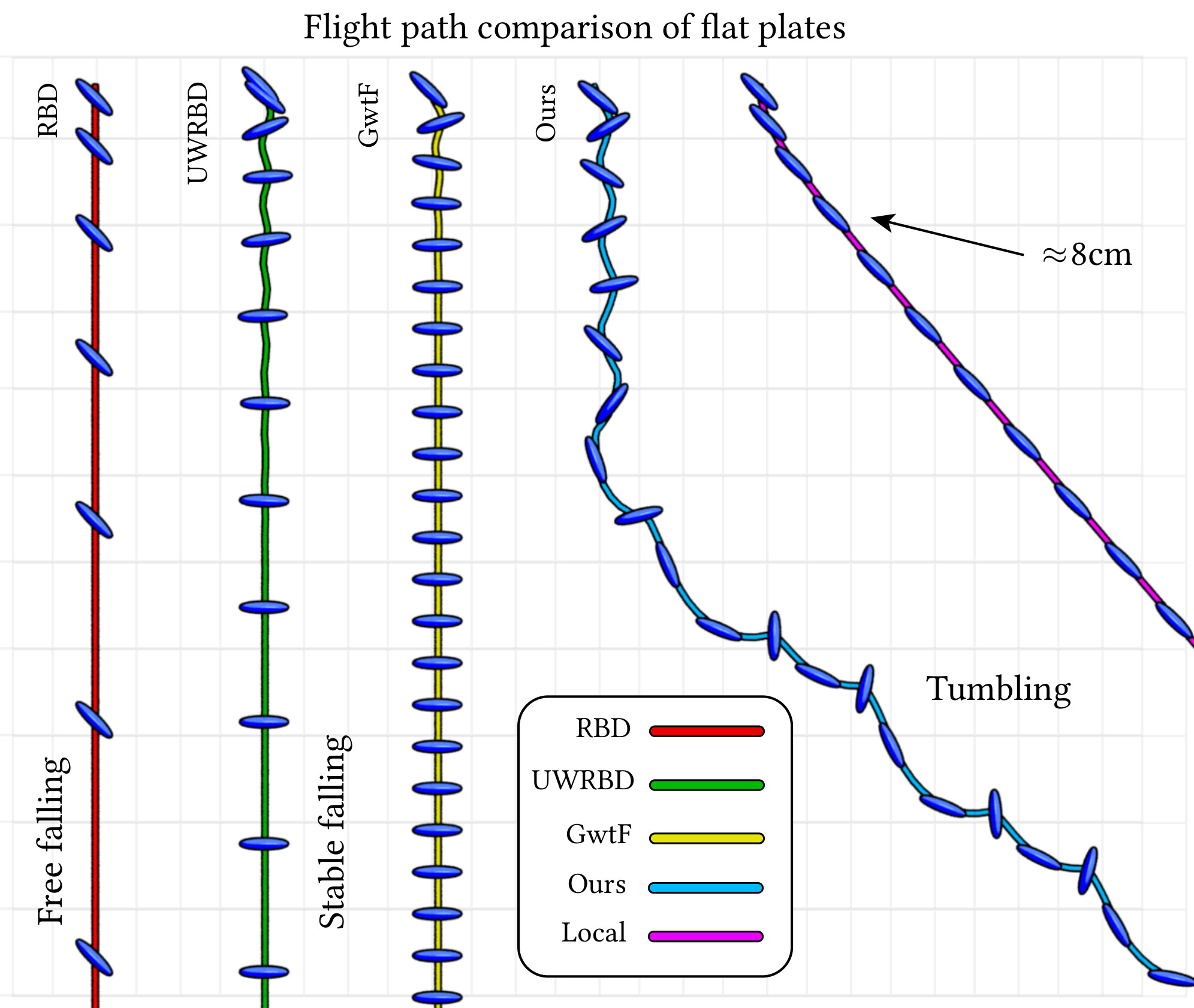}
    \caption{Comparison of falling plates (\SI{8}{\centi\meter} width). Both UWRBD {\protect \cite{weismann_underwater_2012}} and GwtF {\protect \cite{soliman_going_2024}} overdamp the lateral motion, leading to unrealistically stable descent. When UWRBD (using $\lambda = 1$) is given a lower linear drag coefficient, the simulation becomes unstable. The local-only model shows no wobbling at all. Our method reproduces the experimentally observed tumbling mode.}
    \label{fig:falling_plates_comparison}
\end{figure}

\begin{figure}[t]
    \centering
    \includegraphics[width=\linewidth]{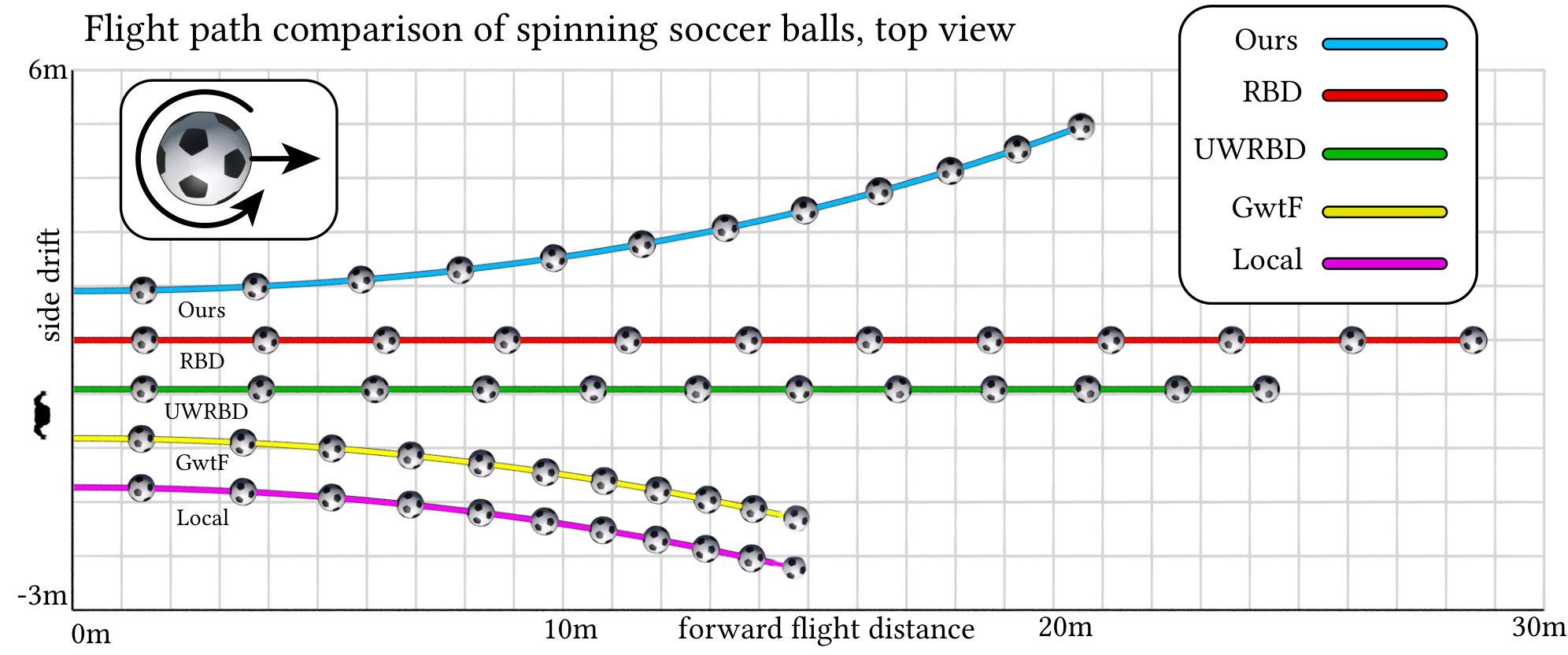}
    \caption{Soccer ball trajectories with identical initial velocity and spin. UWRBD {\protect \cite{weismann_underwater_2012}} neglects lift and produces a nearly straight path with mild drag. GwtF  {\protect \cite{soliman_going_2024}} and the local-only model introduce drift opposite to the Magnus effect. Only our method reproduces the correct lateral deflection.}
    \label{fig:soccer_compare}
\end{figure}

\begin{figure}[t]
    \centering
    \includegraphics[width=\linewidth]{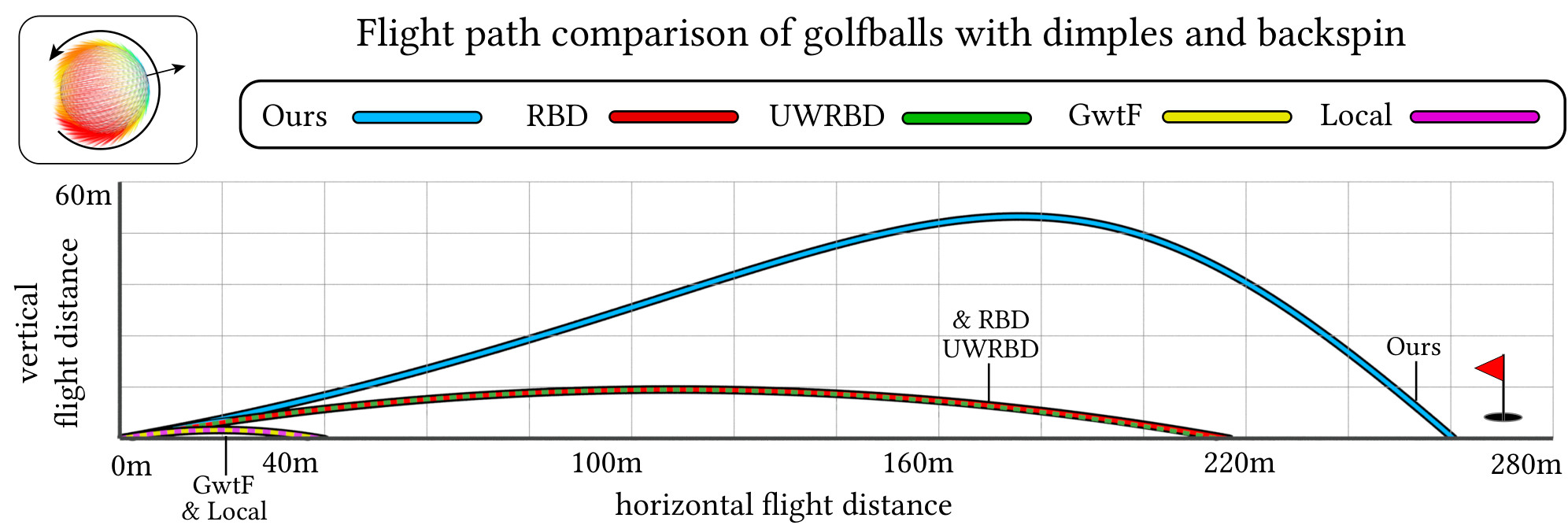}
    \caption{Backspinning dimpled golf ball trajectories. UWRBD {\protect \cite{weismann_underwater_2012}}, GwtF {\protect \cite{soliman_going_2024}}, and local-only force models further shorten the flight relative to rigid body dynamics. Our method recovers the expected increase in range due to backspin {\protect\cite{usga_golf_2023, usga_randa_revised_testing_2023}}.}
    \label{fig:golf_comparison}
\end{figure}

\begin{figure}[t]
    \centering
    \includegraphics[width=\linewidth]{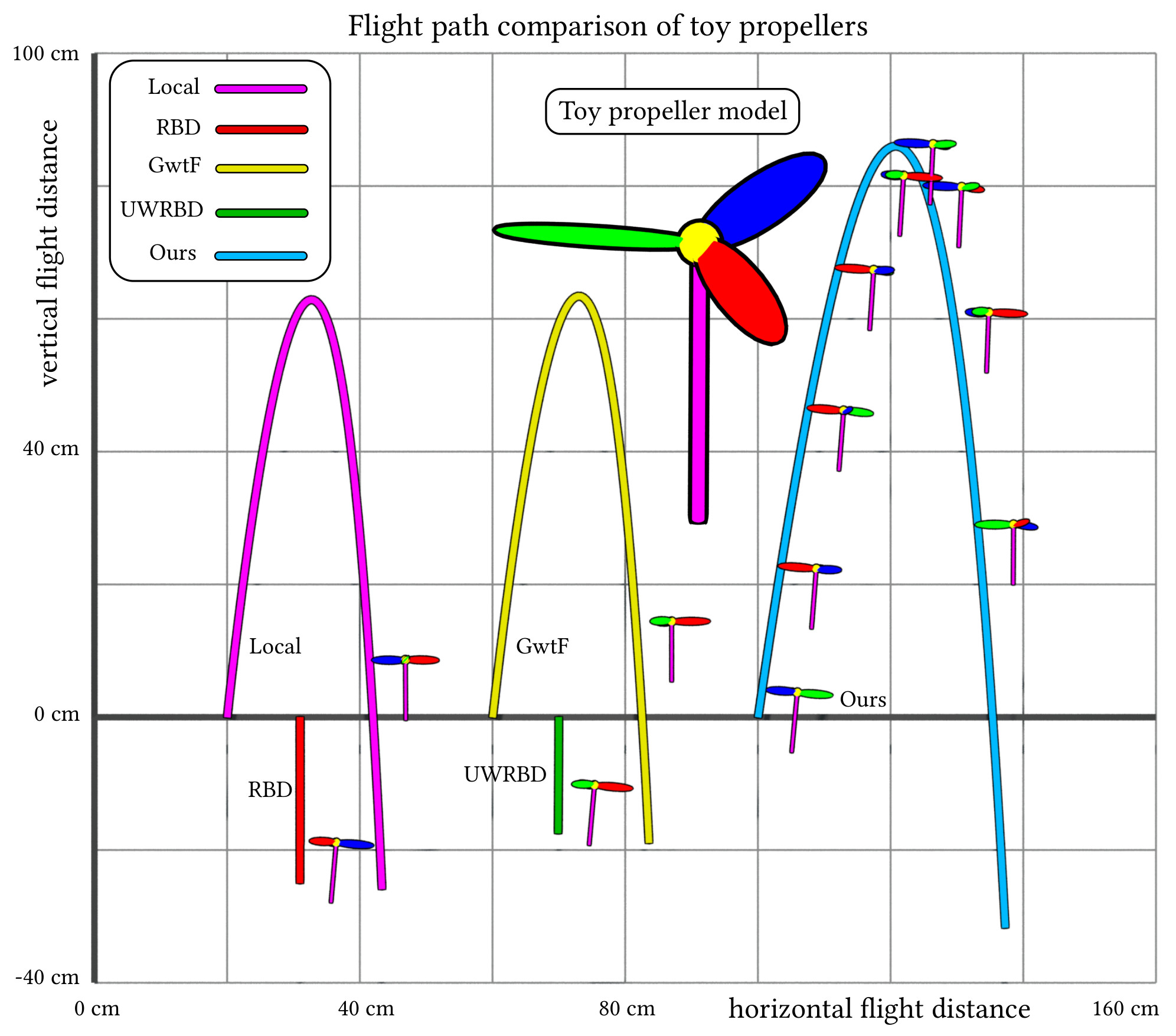}
    \caption{Toy propeller launches with identical initial spin and release angle. UWRBD  {\protect \cite{weismann_underwater_2012}} yields slightly slower descent than rigid body dynamics, while GwtF  {\protect \cite{soliman_going_2024}}, local-only force models, and our method capture the expected lift-induced flight path, although at different scales.}
    \label{fig:toy_comparison}
\end{figure}

\begin{figure}[t]
    \centering
    \includegraphics[width=\linewidth]{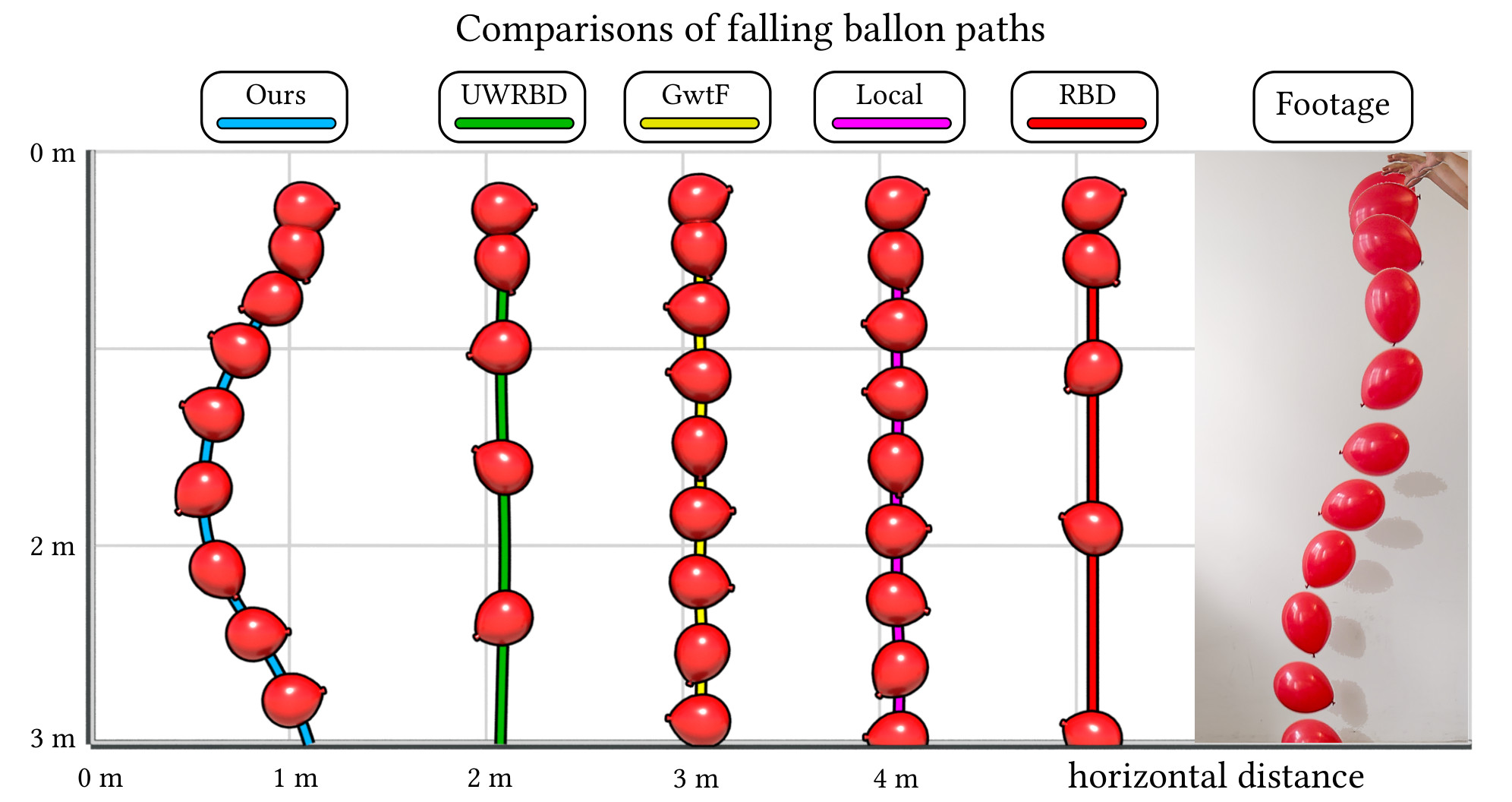}
    \caption{We simulate the dropping of a balloon with non-uniform mass distribution using various methods. Our method is the only one to reproduce the large horizontal motion seen in footage (see supplementary video).}
    \label{fig:balloon_comparison}
\end{figure}

For consistency, all comparisons employ the same inertia tensor evaluation (UWRBD and GwtF use our $\Kfl$, which for GwtF replaces its convex-shape approximation; the local-only model and classical RBD use $\K = \Kb$) and identical numerical integration (RK4). This ensures that differences arise only from the respective fluid force formulations. Figures~\ref{fig:falling_plates_comparison}, \ref{fig:soccer_compare}, \ref{fig:golf_comparison}, \ref{fig:toy_comparison} and \ref{fig:balloon_comparison} illustrate the resulting behaviors across representative experiments.

Overall, UWRBD lacks a lift model and suffers from its linear drag formulation: small parameter choices lead to instabilities, while large values can overdampen the results. They focus on underwater settings to mitigate these problems. GwtF and local-only methods include both lift and drag, producing more stable trajectories, but their choice of forces overdampens the system and suppresses key aerodynamic effects. In particular, these models fail to reproduce the Magnus-effect drift in the correct direction. In contrast, our formulation derives lift and drag implicitly from surface pressures, yielding stable simulations that capture the correct dynamic behaviors without artificial damping.

\begin{table*}[t]
\centering
\caption{Computation times for preprocessing and simulation loop steps of our algorithm when run on an i9-14900KF CPU for all our experiments. We also report the baseline RBD runtime (without fluid interaction), the relative computational overhead and relevant parameters. Fast-spinning objects require significantly more substeps to remain stable.}
\label{tab:performance}
\rowcolors{2}{gray!10}{white}
\begin{tabular}{lcccccccccc}
\toprule
Experiment & plate & soccer & basket & golf & golf spin & football & bunny & toy & balloon & paper \\
Figure     & Fig.\ref{fig:falling_plates} & Fig.\ref{fig:soccer} & Fig.\ref{fig:basketball} & Fig.\ref{fig:golf_ball} & Fig.\ref{fig:golf_ball} & Fig.\ref{fig:football} & Fig.\ref{fig:under_water} & Fig.\ref{fig:toy_propeller} & Fig.\ref{fig:balloon} & Fig.\ref{fig:paper_copter} \\
\midrule
Vertices $\#V \in \mathbb{N}$ & 450 & 252 & 252 & 252 & 252 & 392 & 468 & 526 & 252 & 1347 \\
Faces $\#F \in \mathbb{N}$ & 896 & 500 & 500 & 500 & 500 & 780 & 932 & 1048 & 500 & 2690 \\
Separation angle $\alpha \in [\frac{\pi}{2},\pi]$ & $\frac{\pi}{2}$ & $\frac{\pi}{2}$ & 1.2\,$\frac{\pi}{2}$ & $\frac{\pi}{2},1.3\frac{\pi}{2}$ & $\frac{\pi}{2},1.3\frac{\pi}{2}$ & 1.3\,$\frac{\pi}{2}$ & 1.1\,$\frac{\pi}{2}$ & $\frac{\pi}{2}$ & $\frac{\pi}{2}$ & $\frac{\pi}{2}$ \\
Preprocessing time (ms) & 532.1 & 92.86 & 93.5 & 96.03 & 93.01 & 352.02 & 588.98 & 1.15\,s & 111.4 & 5.08\,s \\
Step time (ms) & 2.58 & 30.22 & 42.97 & 2.05 & 79.07 & 47.98 & 2.73 & 129.75 & 2.02 & 5.44 \\
Step time classic RBD (ms) & 0.51 & 5.85 & 8.47 & 0.59 & 15.9 & 7.89 & 0.56 & 17.19 & 0.56 & 0.59 \\
Overhead factor & $\times5.07$ & $\times5.16$ & $\times5.07$ & $\times3.49$ & $\times4.97$ & $\times6.08$ & $\times4.91$ & $\times7.55$ & $\times3.59$ & $\times9.25$ \\
Time step (s) & 0.03 & 0.03 & 0.03 & 0.02 & 0.02 & 0.03 & 0.03 & 0.03 & 0.03 & 0.01 \\
Max num.\ substeps & 1 & 17 & 24 & 1 & 44 & 21 & 1 & 50 & 1 & 1 \\
\bottomrule
\end{tabular}
\end{table*}

\subsection{Performance}

All experiments are conducted on an i9-14900KF CPU. Performance results are summarized in \tabref{tab:performance}.

\textit{Preprocessing.} The fluid inertia tensor $\Kfl$ is computed by solving six BIEs, each involving the same dense $\#V \times \#V$ linear system. This $\mathcal{O}(\#V^3)$ step constitutes the primary computational bottleneck, but is executed only once per mesh. Preprocessing times are reported in \tabref{tab:performance}.

\textit{Simulation loop.} During simulation, each timestep involves a force evaluation and a single RBD integrator step. The equations of motion operate on six-dimensional state vectors, with the exception of the dynamic pressure force, which requires a parallelizable per-face sum. The per-step timings scale linearly with $\#F$ and are also shown in \tabref{tab:performance}. We report baseline RBD runtimes to quantify the overhead factor introduced by our model.

\subsection{Limitations and future work}\label{sec:limitation}

\begin{figure}[H]
  \centering
  \setlength{\tabcolsep}{3pt}
\renewcommand{\arraystretch}{0.5}
\newlength{\cfdbunnyw}\setlength{\cfdbunnyw}{\dimexpr\linewidth-2\tabcolsep\relax}
\begin{tabular}{c}
  \makebox[\cfdbunnyw]{%
        \makebox[0.2500\cfdbunnyw]{\small Slice}%
        \makebox[0.2500\cfdbunnyw]{\small CFD}%
        \makebox[0.2500\cfdbunnyw]{\small Potential}%
        \makebox[0.2500\cfdbunnyw]{\small Diff}%
  } \\[2pt]
  \includegraphics[width=\cfdbunnyw]{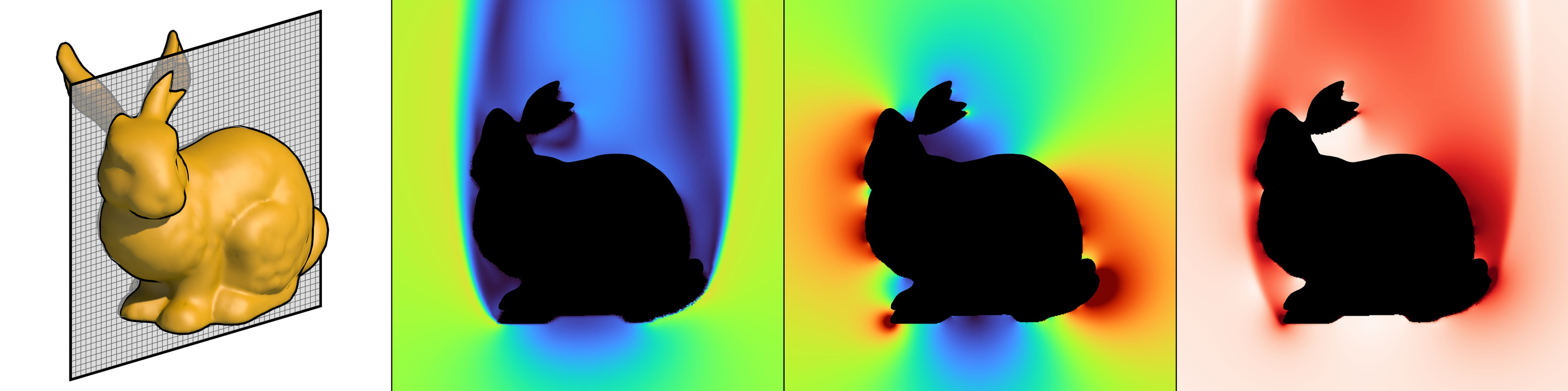} \\[4pt]
  \includegraphics[width=\cfdbunnyw]{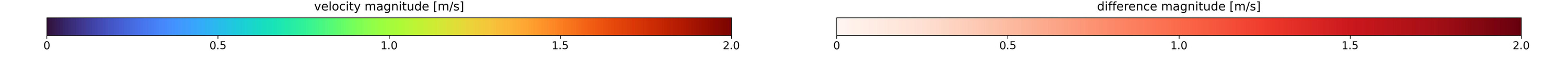} \\
\end{tabular}

  \caption{Surface velocity magnitude slices for the Stanford bunny falling at $v = 1$\,m/s ($\mathrm{Re} \approx 10{,}000$): OpenFOAM~\cite{jasak2007openfoam} RANS, the potential-flow estimate and their absolute difference, highlighting how the current flow separation criterion struggles to capture the real separation. Better flow separation models are needed to capture such shapes more accurately.}
  \label{fig:cfd_bunny}
\end{figure}

We identify a number of limitations of our method. The piecewise linear hat function discretization used to represent the surface potential $\phi$ can under-resolve flow features near sharp edges or thin wings where the potential varies strongly. This is visible in our paper-copter example, where a slightly thickened proxy reproduces the qualitative autorotation but not the descent rate or detailed real-world motion. For the same reason, we also do not reproduce the autorotation of a maple seed, whose thin wing without a larger mass at the bottom poses a strong challenge. The BIE-based preprocessing requires a dense matrix solve, making it impractical for very high-resolution meshes. Our formulation assumes a single rigid body in a still, incompressible fluid and does not handle interactions between multiple bodies, deformable shapes and ambient flow (e.g.\ wind or currents). Furthermore, the flow separation angle $\alpha$ is applied as a single global threshold, whereas real flow separation is much more intricate, especially for more complex shapes (\figref{fig:cfd_bunny}). Scenarios dominated by complex turbulence phenomena, such as vortex shedding and Reynolds-number-dependent flow transitions, remain out of reach; the separation angle $\alpha$ serves as a coarse proxy for these effects.

Several directions for future work emerge from these limitations. Better, local estimations of flow separation depending directly on the body's shape, roughness and/or current Reynolds number would likely be of great benefit to this model. Furthermore, our current formulation of the surface slip velocity in \eqnref{eq:slip} could be refined to better capture subtle flow effects. Addressing these limitations could possibly extend the applicability of our theory to include stable gliding and autorotation of winged seeds.

\section{Conclusion}

We presented a lightweight extension to rigid body dynamics that reproduces complex fluid–body interactions without relying on full fluid simulations, heuristic lift and drag coefficients or shape-specialized solvers. Our method rests on two closely related components: a dynamic pressure force driven by the surface slip velocity, which replaces heuristic drag and lift coefficients with a geometry-aware pressure model, and a flow separation criterion that estimates where this pressure is applied, governed by a single flow separation angle $\alpha$ (see \secref{sec:flow separation} for the underlying assumptions).
Our work is the first to reproduce the full spectrum of falling thin plate modes: fluttering, tumbling, chaotic and steady descent, using such a lightweight setup. This phenomenon has previously relied on full Navier–Stokes simulations, specialized solvers, or 2D models with fitted force coefficients~\cite{pesavento_falling_2004,andersen_unsteady_2005}. Our results suggest that many complex effects could be recovered through added mass and surface pressure estimates alone.

Our method integrates into existing rigid body solvers at about $3.5$--$9\times$ the cost of a classical RBD step (\tabref{tab:performance}). We hope that this work motivates better yet still local estimates of flow separation, which would extend the approach to gliding and autorotating shapes such as winged seeds.

\section*{Acknowledgements}
This work was supported in part by the Alexander von Humboldt Foundation through a Feodor Lynen Research Fellowship and the European Research Council (ERC) under the European Union's Horizon 2020 research and innovation program (grant agreement No.\ 101003104, ERC CoG MYCLOTH).  Additional support was provided by SideFX software.

\printbibliography

\clearpage
\appendix

\section{Potential flow and added mass}\label{app:potential flow}

\subsection{Potential flow}

\begin{figure}
\includegraphics[width=\linewidth]{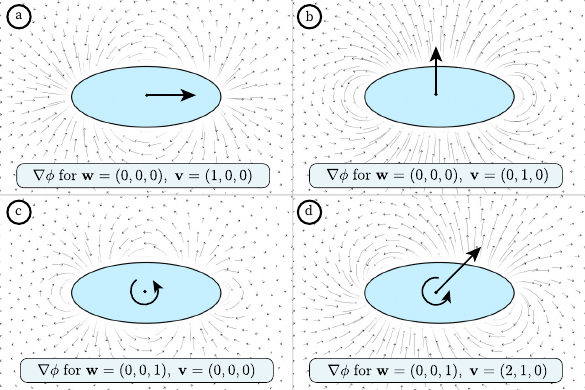}
\caption{Potential flow fields $\uvec = \nabla \phi$ induced by different components of rigid-body motion $(\w, \vvec)$ for an elliptical body (2D-slice view). (a) Horizontal translation induces weak fluid inertia. (b) Vertical translation yields a stronger inertial response. (c) Rotation around the center generates a distinct flow field. (d) General motion results from a linear combination of the six fundamental modes that correspond to the six degrees of freedom of motion (\eqnref{eq:decomposition}).}
\label{fig:flow guide}
\end{figure}

The kinetic energy of the fluid is given by
\begin{equation}
    \Ekinf = \tfrac{1}{2}\,\rhof\int_{\Om} \norm{ \uvec }^2 dV.
\end{equation}
Given $\vvec$ and $\w$, the velocity field $\uvec$ that minimizes $\Ekinf$ is the gradient of the harmonic potential $\phi$ solving the Neumann problem in \eqnref{eq:neumann-problem}. The resulting velocity field $\uvec = \nabla \phi$ represents the incompressible, irrotational flow induced by the rigid body's motion through the fluid. Since the governing Laplace equation and the Neumann boundary condition are both linear in the velocity components $\w, \vvec$, the corresponding solution $\phi$ depends linearly on these inputs. It follows that the general solution can be constructed as a linear superposition of six basis potentials, each corresponding to unit motion in one of the six canonical degrees of freedom (\figref{fig:flow guide}):
\begin{equation}\label{eq:decomposition}
\begin{aligned}
    \phi &= \w_x \phi_{\w_x} + \w_y \phi_{\w_y} + \w_z \phi_{\w_z} + \vvec_x \phi_{\vvec_x} + \vvec_y \phi_{\vvec_y} + \vvec_z \phi_{\vvec_z}\\
    &= \underbrace{[ \phi_{\w_x} \  \phi_{\w_y} \  \phi_{\w_z} \ \phi_{\vvec_x} \  \phi_{\vvec_y} \  \phi_{\vvec_z}] }_{\pot:=}(\w_x,\w_y,\w_z,\vvec_x,\vvec_y,\vvec_z)^{\tp}\\
    &= \pot (\w,\vvec)^{\tp}.
\end{aligned}
\end{equation}
We denote this linear mapping from $(\w, \vvec)$ to the potential field $\phi$ solution of \eqnref{eq:neumann-problem} by $\pot$.

\subsection{Fluid momenta and inertia}

The angular and linear momentum imparted to the fluid by the body's motion (the fluid impulse~\cite{lamb_hydrodynamics_1895}) can be computed explicitly using the divergence theorem. On $S$, the boundary of $\Om$ has the normal $-\n$, since $\n$ points out of the body and into the fluid (and we use $\phi(\z)\rightarrow 0$ for large $\norm{\z}$):
\begin{equation}\label{eq:fluid momenta}
\begin{aligned}
\llvec_f &= \int_\Om \rhof \, \z \times \nabla\phi(\z) \, \dVz = -\rhof \int_S \phi(\z) \, \z \times \n \, \dAz\,, \\
\p_f &= \int_\Om \rhof \, \nabla\phi(\z) \, \dVz = -\rhof \int_S \phi(\z) \, \n \, \dAz\,.
\end{aligned}
\end{equation}
These expressions are linear in $\phi$, and we denote the resulting linear operator $\phi \mapsto (\llvec_f, \p_f)$ as $\phitomom$. Composing this with the linear map $\pot:(\w,\vvec) \mapsto \phi$ yields
\begin{equation}\label{eq:fluid inertia}
\begin{bmatrix}
    \llvec_f \\
    \p_f
\end{bmatrix}
=
\phitomom \, \phi
=
\underbrace{
\phitomom \, \pot
}_{\Kfl :=}
\begin{bmatrix}
    \w \\
    \vvec
\end{bmatrix}.
\end{equation}

\subsection{Discretization}\label{app:discretization}

We discretize the surface potential $\phi$ using continuous, piecewise linear basis functions $h_i$, each defined to take the value $1$ at vertex $\z_i$ and $0$ at all other mesh vertices:
\begin{equation}\label{eq:phihat}
    \phi(\z) = \sum_{i=1}^{\#V} \phi_i h_i(\z).
\end{equation}
In the discrete setting, we treat $\phi$ as an element of $\R^{\#V}$. The surface $S$ is discretized as a union of triangluar faces $f \in F$ with constant normals $\n_i\in\R^3$ during integration. For each vertex $v_i$, we denote by $N_{f,i}$ the set of incident faces sharing that vertex.

\subsection{From potential $\phi$ to momenta $(\llvec,\p)$}\label{app:fluid momentum}

Given a potential field $\phi$ represented in terms of vertex-based linear hat functions $h_i$, we construct the matrix operator $\phitomom \in \R^{6 \times \#V}$ that maps scalar potential to the corresponding fluid-induced angular and linear momenta $(\llvec_f, \p_f)$, as defined in \eqnref{eq:fluid momenta}. Substituting the expansion $\phi(\z) = \sum_i \phi_i h_i(\z)$ into the momentum integrals yields:
\begin{equation}%
\begin{aligned}
\llvec_f &= -\rhof \int_S \phi(\z) \, \z \times \n \, \dAz = \sum_{i=1}^{\#V} \underbrace{\left( -\rhof \int_S h_i(\z) \, \z \times \n \, \dAz \right)}_{ \llvec_{f,i} := } \phi_i, \\
\p_f &= -\rhof \int_S \phi(\z) \, \n \, \dAz = \sum_{i=1}^{\#V} \underbrace{\left( -\rhof \int_S h_i(\z) \, \n \, \dAz \right)}_{ \p_{f,i} := } \phi_i.
\end{aligned}
\end{equation}
The quantities $\llvec_{f,i}$ and $\p_{f,i}$ admit further simplification by exploiting the local support of $h_i$, which is nonzero only on the faces incident to vertex $v_i$:

\begin{equation}\label{eq:fluid momenta expanded}
\begin{aligned}
\llvec_{f,i} &= -\sum_{f_j \in N_{f,i}} \rhof \int_{f_j} h_i(\z) \, \z \, \dAz \times \n_j \\
&= -\sum_{f_j \in N_{f,i}} \frac{1}{12} \rhof A_j \left( 2 \z_{j,i_0} + \z_{j,i_1} + \z_{j,i_2} \right) \times \n_j, \\
\p_{f,i} &= -\sum_{f_j \in N_{f,i}} \rhof \int_{f_j} h_i(\z) \, \dAz \, \n_j 
= -\sum_{f_j \in N_{f,i}} \frac{1}{3} \rhof A_j \, \n_j,
\end{aligned}
\end{equation}
where the indices $(i_0, i_1, i_2)$ denote the local ordering of vertices on face $f_j$ with $\z_{j,i_0} = \z_i$. The matrix $\phitomom$ is then constructed by stacking the row vectors $\llvec_{f,i}$ and $\p_{f,i}$ into a single $6 \times \#V$ matrix:
\begin{equation}
\phitomom = 
\begin{bmatrix}
\llvec_{f,1} & \cdots & \llvec_{f,\#V} \\
\p_{f,1} & \cdots & \p_{f,\#V}
\end{bmatrix}.
\end{equation}

\section{CFD details}\label{app:cfd_setup}

\subsection{Simulation setup}

The reference CFD simulations are run in \textsc{OpenFOAM}~\cite{jasak2007openfoam}~v2512 with the steady-state incompressible solver \texttt{simpleFoam} and the k-$\omega$~SST turbulance model. The fluid is air at $20^\circ$C with kinematic viscosity $\nu = 1.516\times10^{-5}\ \mathrm{m^2/s}$ and density $\rho = 1.204\ \mathrm{kg/m^3}$. The mesh is generated by \texttt{snappyHexMesh} around the body geometry, and the surface velocity is sampled on the body for direct comparison with the potential-flow estimate.

\subsection{Ball: speed variation}\label{sec:cfd_sphere_variation}

\begin{figure}[h]
  \centering
  \setlength{\tabcolsep}{3pt}
\renewcommand{\arraystretch}{0.5}
\newlength{\cfdgridw}\setlength{\cfdgridw}{0.90\linewidth}
\newlength{\cfdgridh}\setlength{\cfdgridh}{1.00923\cfdgridw}
\begin{tabular}{c @{\hspace{4pt}} c}
  &
  \makebox[\cfdgridw]{%
    \makebox[0.3333\cfdgridw]{$\vvec = 0.5$ m/s}%
    \makebox[0.3333\cfdgridw]{$\vvec = 20$ m/s}%
    \makebox[0.3333\cfdgridw]{$\vvec = 40$ m/s}%
  } \\[2pt]
  \rotatebox{90}{%
    \makebox[\cfdgridh]{%
      \makebox[0.32907\cfdgridh]{\centering\small Difference}%
      \hspace{0.01097\cfdgridh}%
      \makebox[0.32907\cfdgridh]{\centering\small Potential}%
      \hspace{0.00183\cfdgridh}%
      \makebox[0.32907\cfdgridh]{\centering\small CFD}%
    }%
  }%
  & \includegraphics[width=\cfdgridw]{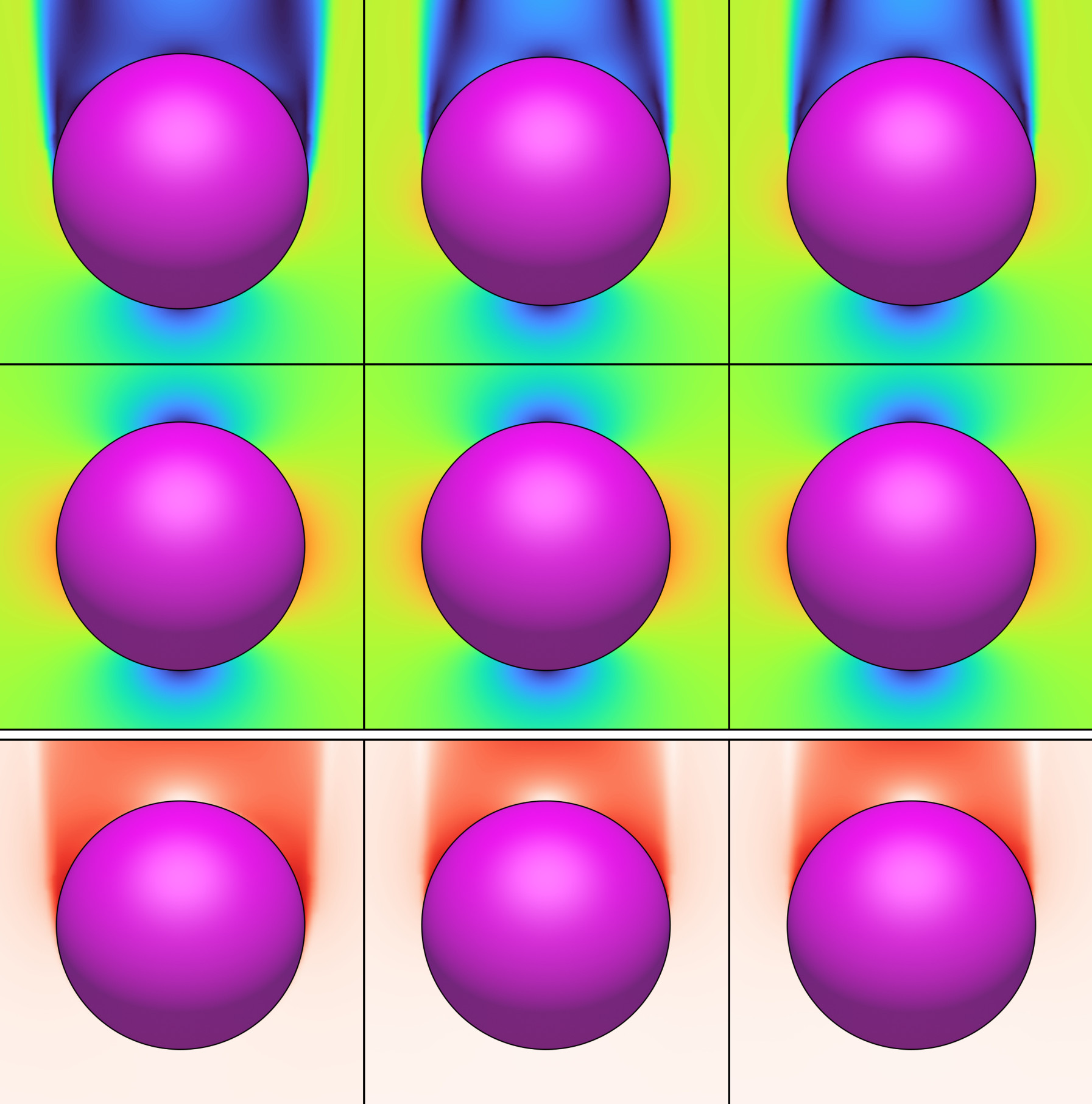} \\[4pt]
  & \includegraphics[width=\cfdgridw]{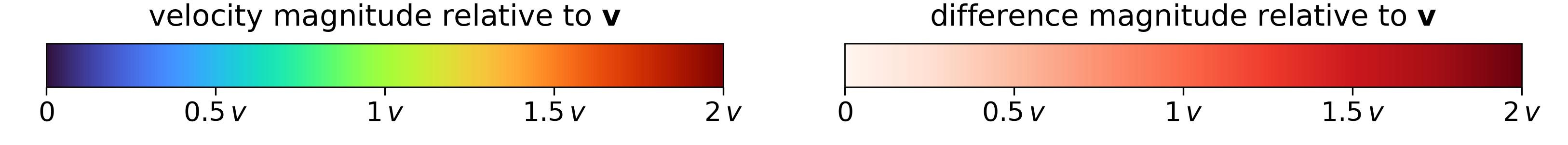} \\
\end{tabular}

    \caption{Surface velocity magnitude for a ball (diameter $10\ \mathrm{cm}$) moving at three different speeds ($v = 0.5$, $20$, $40$ m/s).
    \emph{Top:} \textsc{OpenFOAM}~\cite{jasak2007openfoam} steady RANS reference.
    \emph{Middle:} Potential flow.
    \emph{Bottom:} Absolute difference.
        The side moving towards the fluid matches well with the potential flow. The velocity drop on the side moving away from the fluid is addressed by the separation angle model. The varying speed has little effect on this behavior. Only the speed is varied; the separation angle $\alpha$ is not a parameter of these reference simulations. The consistent behavior across speeds supports using a single, fixed $\alpha$ rather than a speed-dependent one.}
    \label{fig:cfd_sphere_variation}
        
\end{figure}

Figure~\ref{fig:cfd_sphere_variation} compares the surface velocity magnitude of the potential flow against steady RANS \textsc{OpenFOAM}~\cite{jasak2007openfoam} simulations for a sphere at speeds $v = 0.5$, $20$ and $40$ m/s. The side moving towards the fluid matches well. The side moving away from the fluid shows larger differences, which are addressed by setting the slip velocity to zero beyond the flow separation angle $\alpha$. The separation behavior remains consistent across all three speeds, which suggests that a single $\alpha$ applies over a large range of Reynolds numbers. These comparisons motivate the separation region and the omission of $\nabla\phi^{\parallel}$ (\secref{sec:flow separation}); they do not validate $\slip$ itself.

\end{document}